\documentclass[useAMS, usenatbib]{mn2e}
\usepackage{graphicx, rotate}
\usepackage{natbib, lscape}
\usepackage{latexsym, amssymb, verbatim}
\usepackage{eufrak}
\usepackage{amsmath}
\usepackage{epsfig}
\usepackage{times}
\usepackage{rotating}
\usepackage{graphicx}
\usepackage{color}

\newcommand{\apj}{ApJ}
\newcommand{\apjl}{ApJL}
\newcommand{\aap}{A\&A}
\newcommand{\bain}{BAN}
\newcommand{\mnras}{MNRAS}
\newcommand{\rmxaa}{RMxAA}
\newcommand{\aaps}{A\&AS}
\newcommand{\apjs}{ApJS}
\newcommand{\na}{NewA}
\newcommand{\araa}{ARAA}
\newcommand{\aj}{AJ}
\newcommand{\pasa}{PASA}

\title[{\sc StarBench}: The D-type expansion of an H~{\sc ii} region]
{{\sc StarBench}: The D-type expansion of an H~{\sc ii} region}
\author[Bisbas T.~G. et al.]
{Bisbas T.~G.$^1$\thanks{E-mail: ucaptbi@ucl.ac.uk}, Haworth T.~J.$^2$, Williams R.~J.~R.$^3$, Mackey J.$^{4,5}$, Tremblin P.$^{6,7}$, \newauthor Raga A.~C.$^8$,  
 Arthur S.~J.$^9$, Baczynski C.$^{10}$, Dale J.~E.$^{11,12}$,Frostholm T.$^{13}$, Geen S.$^{14}$, \newauthor Haugb{\o}lle T.$^{13}$, Hubber D.$^{11,12}$, Iliev I.~T.$^{15}$,  Kuiper R.$^{16,17}$, Rosdahl J.$^{18}$, Sullivan D.$^{15}$, \newauthor Walch S.$^{5}$, W\"unsch R.$^{19}$\\
$^1$ University College London, 132 Hamstead Road, Kings Cross, NW1 3EE, London, UK\\
$^2$ Institute of Astronomy, University of Cambridge, Madingley Road, Cambridge CB3 0HA, UK\\
$^3$ AWE Aldermaston, Reading, RG7 4PR \\
$^4$ Argelander-Institut f\"ur Astronomie, Auf dem H\"ugel 71, 53121 Bonn, Germany \\
$^5$ I.\ Physikalisches Institut, Universit\"at zu K\"oln, Z\"ulpicher Strasse 77, 50937 K\"oln, Germany \\
$^6$ Astrophysics Group, University of Exeter, EX4 4, QL Exeter, UK \\
$^7$ Maison de la Simulation, CEA-CNRS-INRIA-UPS-UVSQ, USR 3441, CEA Saclay, France \\
$^8$ Instituto de Ciencias Nucleares, Universidad Nacional Aut\'onoma de M\'exico, Mexico \\
$^9$ Centro de Radioastronom\'ia y Astrof\'isica, Universidad Nacional Aut\'onoma de M\'exico, Campus Morelia, Apartado Postal 3-72, 58090, \\ Morelia, Michoac\'an, M\'exico \\
$^{10}$ Zentrum f\"ur Astronomie der Universit\"at Heidelberg, Institut f\"ur Theoretische Astrophysik, Albert-Ueberle-Str. 2, 69120 Heidelberg, Germany \\
$^{11}$ University Observatory, Ludwig-Maximilians-University Munich, Scheinerstr.1, D-81679 Munich, Germany \\
$^{12}$ Excellence Cluster Universe, Boltzmannstr. 2, D-85748 Garching, Germany\\
$^{13}$ Centre for Star and Planet Formation, Natural History Museum of Denmark and Niels Bohr Institute, University of Copenhagen, {\O}ster Voldgade 5-7, \\DK-1350 Copenhagen, Denmark \\
$^{14}$ Laboratoire AIM, Paris-Saclay, CEA/IRFU/SAp - CNRS - Universit\'e Paris Diderot, F-91191 Gif-sur-Yvette Cedex, France\\
$^{15}$ Department of Physics and Astronomy, University of Sussex, Falmer, Brighton BN1 9QH, UK \\
$^{16}$ Institute of Astronomy and Astrophysics, Auf der Morgenstelle 10, D-72076 T\"ubingen, Germany\\
$^{17}$ Max-Planck-Institut f\"ur Astronomie, K\"onigstuhl 17, D-69117 Heidelberg, Germany \\
$^{18}$ Leiden Observatory, Leiden University, PO Box 9513, NL-2300 RA Leiden, the Netherlands \\
$^{19}$ Astronomical Institute, The Czech Academy of Sciences, Bo\v{c}n\'\i ~II 1401, 14131 Prague, Czech Republic}
\date{Accepted , Received ; in original form \today}

\begin{document}

\maketitle

\begin{abstract}
{\sc StarBench} is a project focused on benchmarking and validating different star-formation and stellar feedback codes. In this first {\sc StarBench} paper we perform a comparison study of the D-type expansion of an H~{\sc ii} region. The aim of this work is to understand the differences observed between the twelve participating numerical codes against the various analytical expressions examining the D-type phase of H~{\sc ii} region expansion. To do this, we propose two well-defined tests which are tackled by 1D and 3D grid- and SPH- based codes. The first test examines the `early phase' D-type scenario during which the mechanical pressure driving the expansion is significantly larger than the thermal pressure of the neutral medium. The second test examines the `late phase' D-type scenario during which the system relaxes to pressure equilibrium with the external medium. Although they are mutually in excellent agreement, all twelve participating codes follow a modified expansion law that deviates significantly from the classical Spitzer solution in both scenarios. We present a semi-empirical formula combining the two different solutions appropriate to both early and late phases that agrees with high-resolution simulations to $\lesssim2\%$. This formula provides a much better benchmark solution for code validation than the Spitzer solution. The present comparison has validated the participating codes and through this project we provide a dataset for calibrating the treatment of ionizing radiation hydrodynamics codes. 
\end{abstract}


\section{Introduction}%

Computer simulations play an increasingly important role in Astrophysics. They allow researchers to apply theoretical understanding to complex systems beyond the capability of analytic calculation, and also provide validation for the physical correctness of proposed theoretical models. However, computational models have their own sources of error. It is therefore important to validate their accuracy, where possible, if their results are to have any credence. Different sources of error require different validation techniques, but use of code inter-comparisons and resolution studies for well-defined test problems is particularly powerful in this regard. Previous such comparisons have included photoionization modelling \citep[][]{Pequ01}, PDR code comparison \citep[][]{Roll07}, and cosmological code comparison \citep[][]{Ilie06, Ilie09}.

As well as the standard tests which are widely applied to computational hydrodynamics codes, such as Sod shock tubes \citep{Sod78} and the Sedov-Taylor point explosion \citep{Sedo59,Tayl50}, it is important to develop test problems more specific to the domain of study, which test broader physics couplings. The {\sc StarBench} project seeks to develop a collection of well-understood validation cases for the modelling of star-forming regions. To date, two workshops have been held (Exeter, April 2013; Bonn, September 2014)\footnote{{\sc StarBench-I}: http://www.astro.ex.ac.uk/people/haworth/\\workshop\_bench/index.html \\{\sc StarBench-II}: https://www.astro.uni-bonn.de/sb-ii/index.html}, in which groups specializing in numerical star formation have gathered to set the basis of standard and well-defined tests and explore the differences observed between their codes. The present paper reports the results of the first test investigated by the {\sc StarBench} project: the D-type expansion of an H~{\sc ii} region.

The problem of the expansion of an H~{\sc ii} region into a uniform neutral medium has a long history in theoretical and numerical studies of the interstellar medium (ISM). The ionizing photons ($h\nu > 13.6$~eV) produced by a hot, massive star ionize and heat a volume of gas around the star, which then begins to expand supersonically into the surrounding medium due to the overpressure. The photoionized region is bounded by an ionization front, which separates the ionized from the neutral gas. \citet{Kahn54} and \citet{Axfo61} showed analytically that the ionization front relatively quickly switches from R-type to D-type and drives a strong shock into the neutral ISM. This behaviour was verified by early numerical investigations of H~{\sc ii} region expansion conducted by \citet{Math65} and \cite{Lask66}. These early analytic and numerical calculations are reviewed by \citet{Math69}, and subsequent developments, including the first multi-dimensional simulations in uniform and non-uniform media, are reviewed by \citet{York86}. An analytic solution for the expansion rate of the H~{\sc ii} region during the D-type phase as a function of time was derived by \citet{Spitz78}, which is presented more lucidly in the book of \citet{Dyso80}. This has become the standard solution with which numerical results are compared.

Most modern code developers use the spherical expansion of an H~{\sc ii} region as a standard test problem to validate their coupled photoionization and (magneto-)hydrodynamics algorithms \citep[e.g.][]{Mell06a,Krum07,Mack10,Mack11,Tremb12a,Raga12}. This problem involves a complicated combination of fluid dynamics, radiative transfer, microphysical heating, cooling, ionization and recombination, and differences between results obtained with different codes and algorithms have not so far been addressed in any depth in the literature. There is therefore a pressing need of setting up benchmarking tests to examine the behaviour of the different codes in reproducing known analytical expressions.

Although the \cite{Spitz78} analytical solution is widely used for code testing, alternative analytical solutions for the expansion of an H~{\sc ii} region, applicable at different stages of the evolution and based on different assumptions, have been developed more recently, such as those of \citet{Hoso06} and \citet{Raga12,Raga12b}. It is an aim of the present paper to determine which, if any, of the available analytical solutions is most suitable, and thus to provide a benchmark useful for any other radiation hydrodynamics code. For instance, in the first attempts of the present test, several discrepancies between the codes have been identified resulting from bugs or incorrect assumptions which have all been fixed throughout this work. This demonstrates the value of such a comparison in improving confidence in the accuracy of these codes when applied to more complex scenarios.

The interest of the authors of the present paper in establishing a reliable benchmark test for expanding H~{\sc ii} regions stems from the large body of work, much of it done in the last decade, on the influence of expanding H~{\sc ii} regions at and below giant molecular cloud (GMC) scales, intended to explore explicitly the role of photoionisation in the evolution and destruction/dispersal of GMCs and the regulation of star-formation rates and efficiencies. H~{\sc ii} region feedback has been invoked both as a trigger of star formation \citep[e.g.][]{Elme77}, and as the main agent responsible for terminating it \citep[e.g.][]{McKe84,Matz02}. Many simulations have sought to model one or both of these scenarios in order to establish whether the influence of photoionising feedback on the galactic star-formation process is overall positive or negative.

For example, the gravitational instability of the dense shells driven by H~{\sc ii} regions, often referred to as the `collect and collapse’ process has been modelled by \citet{Elme77,Whit94a,Dale07}, who show that, in principle, it should be an efficient means of triggering star formation in even smooth media. Other authors have instead studied the influence of substructure encountered by the radiation field, either in terms of well–defined isolated clumps and the radiation-driven implosion model \citep{Bert89,Lefl94,Mell98,Kess03,Henn09,Mack11,Bisb11,Hawo12}, or of more generalised fractal substructure \citep{Cole84,LeVe02,Walc11,Walc12,Walc13}.

The negative effects of H~{\sc ii} region expansion on clouds already in a dynamical state have also received much attention. \citet{Pete10} and \citet{Pete11} model H~{\sc ii} regions driven by roughly centrally-located massive stars on rotating GMCs, finding them rather ineffective at preventing the formation of stars. The general problem of the interaction of expanding H~{\sc ii} regions with turbulent media has been modelled by \citet{Mell06a,Grit09,Arth11,Tremb12b,Dale12,Coli13,Bone15}, most of whom conclude that H~{\sc ii} regions are able to drive turbulence, and are destructive to GMCs of $\sim10^4\,{\rm M}_\odot$.

This paper is organized as follows. In Section \ref{sec:theory} we present a theoretical background of the D-type expansion of an H~{\sc ii} region and discuss the different analytical solutions to this problem. In Section \ref{sec:codes} we present the numerical codes participating in this test, highlighting in brief the methods used to propagate the ionizing radiation. In Section \ref{sec:test} we present the initial conditions of the benchmarking test and in Section \ref{sec:results} we show the results. We summarize and conclude in Section \ref{sec:sumcon}.

\section{Theoretical background}%
\label{sec:theory}

A hot, massive star emits large numbers of extreme ultraviolet (EUV) photon with energy $h\nu$ larger than the ionization potential of hydrogen (13.6eV). The ionizing photons interact with the surrounding neutral medium and ionize a volume of gas within which the rate of ionizations is nearly balanced by recombinations with the remaining small flux ionizing the neutral material passing through its surface. The kinetic energy of the ejected photoelectrons heats the gas within the photoionized volume by collisions but efficient cooling due to collisionally excited line radiation of metal ions such as doubly ionized oxygen means that temperatures inside H~{\sc ii} regions are approximately uniform and typically $\sim 10^4$~K \citep[see][for details]{Oste89}. In the analysis that follows, we will ignore the detailed physical processes and make a series of simplifying assumptions.

Consider a spherically symmetric cloud of radius $R_{\rm cl}$, of total mass $M_{\rm cl}$ and of uniform density $\rho_{\rm cl}$ consisting of pure atomic hydrogen with a uniform temperature $T_{\rm cl}$. Let us place a radiation source at the centre of the cloud which defines the origin of a Cartesian coordinate system. The source emits $\dot{\cal N}_{_{\rm LyC}}$ Lyman continuum ionizing photons per unit time and we assume that the ionizing photons are monochromatic with energy $h\nu=13.6~{\rm eV}$. We assume a photoionization cross-section of $\bar{\sigma}=6.3\times10^{-18}\,{\rm cm}^2$ and we use the recombination coefficient, $\alpha_{_{\rm B}}$, into excited stages only by invoking the \emph{on-the-spot} approximation \citep[][known as the `case-B' recombination coefficient]{Oste89}. Assuming an isothermal H~{\sc ii} region at $T_{\rm i}=10^4\,{\rm K}$ the case-B recombination coefficient is taken to be $\alpha_{_{\rm B}}\simeq2.7\times10^{-13}\,{\rm cm}^{3}\,{\rm s}^{-1}$. Hereafter, the indices `i' and `o' shall denote the ionized and the neutral medium, respectively.

\citet{Strom39} was the first to show that the transition zone between the ionized gas and the surrounding neutral medium occurs over a very short distance compared to the dimensions of the H~{\sc ii} region. This transition zone is the {\emph{ionization front}} and we can treat it as a sharp discontinuity. The distance over which the degree of ionization changes from 90\% to 10\% is given by
\begin{eqnarray}\label{eqn.whitworth00}
\Delta R_{\rm St} \simeq \frac{20m_{\rm p}}{\rho_{\rm o}\bar{\sigma}}\simeq1.72\times10^{-4}\,{\rm pc}\left(\frac{\rho_{\rm o}}{10^{-20}\,{\rm g}\,{\rm cm}^{-3}}\right)^{-1}\,,
\end{eqnarray}
\citep[e.g.][]{Whit00} where $m_{\rm p}$ represents the proton mass.

\citet{Kahn54} studied in detail the propagation of an ionization front into a neutral medium when an ionizing star suddenly switches on. At early times most of the Lyman continuum photons ionize additional gas beyond the instantaneous position of the ionization front. Thus the H~{\sc ii} region expands rapidly at highly supersonic speed relative to the sound speed of the ionized gas. In this first phase the ionization front is called {\emph{R-type}} (R=Rarefied). As shown by \citet{Strom39}, the $\dot{\cal N}_{_{\rm LyC}}$ photons emitted by the central source will ionize a spherical region of radius
\begin{eqnarray}\label{eqn.stromgren}
R_{\rm St}=\left(\frac{3{\dot{\cal N}_{_{\rm LyC}}m_{\rm p}^2}}{4\pi\alpha_{_{\rm B}}\rho_{\rm o}^2}\right)^{1/3}\,,
\end{eqnarray}
where $\rho_{\rm o}=\rho_{\rm cl}$ and $m_{\rm p}$ is the proton mass.
Hereafter, we refer to the spherical region of radius $R_{\rm St}$ as the {\it initial Str{\"o}mgren sphere} and the radius $R_{\rm St}$ as the {\it initial Str{\"o}mgren radius}. The R-type phase of expansion terminates once the ionization front reaches this initial Str{\"o}mgren radius. The timescale for this first phase is of order the recombination time
\begin{eqnarray}\label{eqn.rtypetimescale}
t_{_{\rm D}}=\frac{m_{\rm p}}{\alpha_{_{\rm B}}\rho_{\rm o}}\simeq19.6\,{\rm yrs}\left(\frac{\rho_{\rm o}}{10^{-20}\,{\rm g}\,{\rm cm}^{-3}}\right)^{-1}\,.
\end{eqnarray}

The large temperature difference between the two regions results in a large difference in thermal pressure, and the H~{\sc ii} region expands. In this second phase, which is called {\emph{D-type}} (D=Dense), the ionization front propagates at subsonic speed relative to the ionized gas but at supersonic speed relative to the neutral gas. Therefore, it is preceded by a {\emph{shock front}} which sweeps up a dense shell of neutral gas. The shell is bounded on its inside by the ionization front and on its outside by the shock front.

Within the ionized region we assume that at all times there is a balance between the ionizing photons produced by the star and the recombination events. Therefore,
\begin{eqnarray}
\dot{\cal N}_{_{\rm LyC}}=\int \frac{\rho_{\rm i} \rho_{\rm e}}{m_{\rm p}m_{\rm e}}\alpha_{\rm B}dV,
\end{eqnarray}
where $\rho_{\rm e}$ and $m_{\rm e}$ are the electron density and mass respectively. If we assume that i) the medium is entirely ionized, ii) $\rho_{\rm e}=\rho_{\rm i}m_{\rm e}/m_{\rm p}$, and iii) that the ionization front has a sharp edge, the above integral leads to
\begin{eqnarray}
\label{eqn:balance}
\dot{\cal N}_{_{\rm LyC}}=\frac{4\pi}{3}\left\{\frac{\rho_{\rm i}(t)}{m_{\rm p}}\right\}^2\alpha_{_{\rm B}}R^3_{_{\rm IF}}(t),
\end{eqnarray}
which, when combined with Eqn. (\ref{eqn.stromgren}), gives
\begin{eqnarray}\label{eqn.rhoi}
\rho_{\rm i}(t)=\rho_{\rm o}\left\{\frac{R_{\rm St}}{R_{_{\rm IF}}(t)}\right\}^{3/2},
\end{eqnarray}
where $R_{_{\rm IF}}$ is the extend of the ionized region. As the H~{\sc ii} region expands, it ionizes more neutral gas and its mass increases in time. The time-dependent mass of the ionized region, $M_{\rm i}(t)$ is given by:
\begin{eqnarray}\label{eqn.mi}
M_{\rm i}(t)=\frac{4\pi}{3}\rho_{\rm o}R_{\rm St}^{3/2}R_{_{\rm IF}}^{3/2}(t)\label{eqn.mion}\,.
\end{eqnarray}

The thermal pressure of the ionized gas in an H~{\sc ii} region (which drives the expansion) matches approximately with the thermal pressure of the neutral gas in the shell between the ionization front and the shock front. 

\subsection{Early time behaviour}
\label{ssec:early}
\subsubsection{Spitzer approximation}
\label{sssec:spitzer}
The first analytical attempt to model the D-type expansion of an H~{\sc ii} region was performed by \citet{Spitz78} which was investigated in more depth by \citet{Dyso80}. As pointed out by \citet{Raga12}, by assuming the thin shell approximation and by equating the pressure of the neutral gas in the shell between the ionization front and the shock front with the ram pressure of the undisturbed neutral gas as it is swept up by the shock front we obtain:
%
\begin{eqnarray}\label{eqn.raga}
\frac{1}{c_{\rm i}}\frac{d R_{\rm Sp}(t)}{dt}=\left\{\frac{R_{\rm St}}{R_{\rm Sp}(t)}\right\}^{3/4}-\frac{\mu_{\rm i} T_{\rm o}}{\mu_{\rm o} T_{\rm i}}\left\{\frac{R_{\rm St}}{R_{\rm Sp}(t)}\right\}^{-3/4}\, ,
\end{eqnarray}
where $\mu$ is the mean molecular weight of the gas. The above equation shall be referred to as `Raga-I' throughout the present paper.

The ratio ${\mu_{\rm i} T_{\rm o}}/{\mu_{\rm o} T_{\rm i}}$ found on the right-hand-side of Eqn. (\ref{eqn.raga}) is generally small ($\sim 1/200$). At early times the term $\frac{\mu_{\rm i} T_{\rm o}}{\mu_{\rm o} T_{\rm i}}\left\{\frac{R_{\rm St}}{R_{\rm Sp}(t)}\right\}^{-3/4}$ can be neglected, therefore Eqn. (\ref{eqn.raga}) leads to the so-called Spitzer solution:
\begin{eqnarray}\label{eqn.Spitzer}
R_{\rm Sp}(t)=R_{\rm St}\left(1+\frac{7}{4}\frac{c_{\rm i}t}{R_{\rm St}}\right)^{4/7}\,.
\end{eqnarray}

\subsubsection{Hosokawa-Inutsuka approximation}
\label{sssec:hi}
A different approach describing the expansion of an H~{\sc ii} region was provided by \citet{Hoso06} who derived the position of the ionization front in time directly from the equation of motion of the expanding shell. Independently, \citet{Raga12b} argued that the differential Eqn. (\ref{eqn.raga}) does not incorporate the inertia of the shocked neutral gas which is created during the expansion of the H~{\sc ii} region. To include the inertia, we may write the equation of motion of the mass $M$ within the shocked shell as
\begin{eqnarray}\label{eqn.raga2}
\frac{d}{dt}(M\dot R_{\rm HI})=4\pi R_{\rm HI}^2(P_{\rm i}-P_{\rm o}),
\end{eqnarray}
where $P$ is the thermal pressure of the gas and $R_{\rm HI}$ is the position of the shell as modelled by \citet[][`HI']{Hoso06}. The term $P_{\rm i}-P_{\rm o}$ corresponds to the net thermal pressure taking into account the pressure acting from the neutral gas onto the ionized gas. Using Eqn. (\ref{eqn.rhoi}) and where $R_{\rm IF}(t)$ represents now the position of the shocked shell $R_{\rm HI}$, we obtain the second-order non-linear differential equation
\begin{eqnarray}\label{eqn.raga2diff}
\ddot{R}_{\rm HI}+\left(\frac{3}{R_{\rm HI}}\right)\dot{R}_{\rm HI}^2=\frac{3R_{\rm St}^{3/2}c_{\rm i}^2}{R_{\rm HI}^{5/2}}-\frac{3c_{\rm o}^2}{R_{\rm HI}}\,.
\end{eqnarray}
The above equation shall be referred to as `Raga-II' throughout the present paper.
Equation (\ref{eqn.raga2diff}) represents the equation of motion of the expanding shell when its inertia is taken into account. 
Solving the above leads to
\begin{eqnarray}\label{eqn.bernsol}
\dot{R}_{\rm HI}(t)=c_{\rm i}\sqrt{\frac{4}{3}\frac{R_{\rm St}^{3/2}}{R_{\rm HI}^{3/2}(t)}-\frac{\mu_{\rm i}T_{\rm o}}{2\mu_{\rm o}{T_{\rm i}}}}.
\end{eqnarray}
At early times we may neglect the ${\mu_{\rm i} T_{\rm o}}/{\mu_{\rm o} T_{\rm i}}$ term. This corresponds to the assumption of $P_{\rm o}=0$ in Eqn. (\ref{eqn.raga2}). Equation \ref{eqn.bernsol} has therefore the analytical solution
\begin{eqnarray}
\label{eqn:HI}
R_{\rm HI}(t)=R_{\rm St}\left(1+\frac{7}{4}\sqrt{\frac{4}{3}}\frac{c_{\rm i}t}{R_{\rm St}}\right)^{4/7}\,.
\end{eqnarray}
This equation was first presented by \citet{Hoso06} and it differs from the known Spitzer expression (Eqn.~\ref{eqn.Spitzer}) by a $\sqrt{4/3}$ factor. This factor arises from our inclusion of the inertia of the shocked gas due to its own movement leading to a slightly faster expansion than that obtained from Eqn. (\ref{eqn.Spitzer}).

\subsection{Late time behaviour}
\label{ssec:late}

\subsubsection{Raga's extension of Spitzer (Raga-I)}
\label{sssec:ragai}
At later times the term $\frac{\mu_{\rm i} T_{\rm o}}{\mu_{\rm o} T_{\rm i}}\left\{\frac{R_{\rm St}}{R_{\rm Sp}(t)}\right\}^{-3/4}$ increases and eventually the H~{\sc ii} region stagnates at $t=t_{_{\rm STAG}}$ which is defined by $\dot{R}_{\rm Sp}(t_{_{\rm STAG}})=0$. By this time the H~{\sc ii} region is in pressure equilibrium, thus it does not expand further. The stagnation radius is (from Eqn.\ref{eqn.raga})
\begin{eqnarray}\label{eqn.rstag}
R_{_{\rm STAG}}=\left(\frac{c_{\rm i}}{c_{\rm o}}\right)^{4/3}R_{\rm St},
\end{eqnarray}
the density of the ionized gas is (from Eqn.\ref{eqn.rhoi})
\begin{eqnarray}
\rho_{\rm i}=\rho_{\rm o}\left(\frac{c_{\rm o}}{c_{\rm i}}\right)^2,
\end{eqnarray}
and the total ionized mass is (from Eqn.\ref{eqn.mi})
\begin{eqnarray}\label{eqn.massstag}
M_{\rm i}(t_{_{\rm STAG}})=\frac{4\pi}{3}R_{\rm St}^3\rho_{\rm o}\left(\frac{c_{\rm i}}{c_{\rm o}}\right)^2\,.
\end{eqnarray}

\subsubsection{Raga's extension of Hosokawa-Inutsuka (Raga-II)}
\label{sssec:ragaii}

At later times $R_{\rm HI}$ becomes large, and therefore the two terms contained in the square root of Eqn. (\ref{eqn.bernsol}) become comparable. Eventually at $t=t_{_{\rm STAG}}$ in which $\dot{R}_{\rm HI}(t_{_{\rm STAG}})=0$ we obtain the stagnation radius
\begin{eqnarray}
\label{eqn.rstagii}
R_{_{\rm STAG}}=R_{_{\rm St}}\left(\frac{8}{3}\right)^{2/3}\left(\frac{c_{\rm i}}{c_{\rm o}}\right)^{4/3},
\end{eqnarray}
the density of the ionized medium
\begin{eqnarray}
\rho_{\rm i}=\rho_{\rm o}\left(\frac{8}{3}\right)^{2/3}\left(\frac{c_{\rm o}}{c_{\rm i}}\right)^{4/3},
\end{eqnarray}
and the total ionized mass
\begin{eqnarray}
M_{\rm i}(t_{_{\rm STAG}})=\frac{4\pi}{3}R_{\rm St}^3\rho_{\rm o}\left(\frac{8}{3}\right)^{8/3}\left(\frac{c_{\rm i}}{c_{\rm o}}\right)^{8/3}\,.
\end{eqnarray}

\section{Numerical codes}%
\label{sec:codes}

The methods\footnote{It is not the scope of the present paper to describe each hydrodynamical method. We redirect the reader to the cited works for further details.} which the present numerical codes used to account for hydrodynamics are either grid-based techniques \citep[e.g.][]{Cole84,Roe86,Berg89} or smoothed particle hydrodynamics-based techniques \citep[SPH; e.g.][]{Lucy77,Ging77}. Here we present a brief review of the participating codes and their radiative transfer capabilities. All codes are summarized in Table \ref{tab:allcodes}.

\begin{table*}
\caption{ Participating codes in this project.}
\centering
\label{tab:allcodes}
\begin{tabular}{l l c c c c c c c}
\hline
\hline
          &     &      & \multicolumn{2}{c}{Early Phase} & \multicolumn{2}{c}{Late Phase} & Mesh motion & Mesh geometry$^1$ \\
Code name & Code representative(s) & Kind & 1D & 3D			 & 1D & 3D & & \\
\hline
{\sc aquiline} 		& Williams		& Grid	& X &   & X &   & Eulerian & Uniform AMR radial\\
{\sc capreole-C$^2$-Ray}& Arthur		& Grid 	&   & X &   & X & Eulerian  & Uniform cartesian\\
{\sc flash-fervent} 	& Baczynski		& Grid  &   & X &   & X  & Eulerian & Uniform cartesian\\
{\sc flash-treeray} 	& W{\"u}nsch		& Grid	&   & X &   & X  & Eulerian & Uniform cartesian\\
{\sc glide}		& Williams		& Grid	& X &   & X &   & Lagrangian & Uniform radial\\
{\sc heracles} 		& Tremblin		& Grid	& X &   & X &   & Eulerian & Uniform radial\\
{\sc pion}		& Mackey		& Grid	& X & X & X & X & Eulerian & Uniform radial (1D) cartesian (3D)\\
{\sc ramses-lampray} 	& Haugb{\o}lle \& Frostholm & Grid &   & X &   &  & Eulerian & Uniform cartesian\\
{\sc ramses-rt}		& Geen \& Rosdahl	& Grid  &   & X &   & X & Eulerian & Uniform cartesian\\
{\sc sedna}		& Kuiper		& Grid	& X & X & X & X & Eulerian & Uniform radial\\
{\sc seren} 		& Hubber \& Bisbas 	& SPH 	&   & X &   & X & Lagrangian (SPH) & Equal mass \\
{\sc torus}		& Haworth		& Grid	& X &   & X &   & Eulerian & Uniform radial\\
\hline
\end{tabular}
\begin{center}
$^1$ Refers to the test problem presented in this paper and not to the general capabilities of each code.
\end{center}
\end{table*}

\subsection{{\sc aquiline}, {\sc glide} (R. Williams)}

{\sc aquiline} is a simplified version of {\sc aqualung} \citep{Will00}, restricted to one-dimensional problems.  It has been extended to treat problems in spherical coordinates.  Mesh refinement is implemented by bisection of cells, with simple gradient-based refinement criteria.  It is based on a second-order MUSCL scheme using an exact Riemann solver, as in \citet{Fall91}; some improvements have been made to the robustness and accuracy of the Riemann solution in extreme limits.  It uses a photon-conservative scheme for the interaction of the ionizing radiation with the gas, as discussed in \citet{Will02}, but including the obvious modifications for spherical geometry.

{\sc glide} is a one-dimensional Lagrangian code, based on the same underlying Riemann solver and ionization/radiation solve as {\sc aqualung} \citep{Will02}.  This code uses piecewise-linear reconstruction within mesh cells; however, the interface velocities and pressures calculated by the Riemann solver are used to advance the zone widths, momenta and energies directly, rather than being converted into fluxes.  In the calculations presented, the zones have been taken to have constant initial spatial width.

For both of these codes, the hydrodynamical advance was made using $\gamma = 5/3$; testing has shown that this makes a minimal difference from $\gamma$ close to unity, due the close coupling of the radiation step (at both half- and full-step).

\subsection{{\sc capreole-C$^2$-Ray} (G. Mellema, S.J. Arthur)}
\label{ssec:arth}

The {\sc capreole-C$^2$-Ray} radiation-hydrodynamics code combines the nonrelativistic Roe solver PPM scheme described in \citet{Euld95} with the radiation transport and photoionization code C$^2$-Ray (Conservative Causal ray) code developed by \citet{Mell06b}. This code has been used to model the expansion of H~{\sc ii} regions in turbulent molecular clouds \citep{Mell06a,Medi14} and for these applications the various contributions to the heating and cooling in the photoionized and neutral gas are treated by analytical fits, as described in \citet{Henn09}. For the present benchmark test, a simple non-equilibrium prescription for thermal balance has been included, which ensures a temperature of $10^4\,{\rm K}$ in the ionized gas and $10^2\,{\rm K}$ or $10^3\,{\rm K}$, as required, in the neutral gas. The calculations are all performed on a fixed, uniform Cartesian grid in three dimensions and limited parallelisation using Open-MP has been implemented.

\subsection{{\sc flash-fervent} (C. Baczynski)}

{\sc fervent} \citep{Bacz15} is a multi-source, Cartesian, radiative transfer code module implemented in the magnetohydrodynamical grid code {\sc flash 4} \citep{Fryx00, Dube08}. It is based on an adaptively split ray tracing scheme introduced in \cite{Wise11}. Rays are initially cast from a spherically uniform distribution based on the {\sc healpix} decomposition \citep{Gors05}. They intersect and traverse the domain until a splitting criterion is fulfilled, based on the apparent pixel size in comparison to the cell face area. In this way a full sampling of the mesh domain is guaranteed. Each cell of the mesh is intersected by multiple non-aligned rays, which ensures that artifacts from the underlying Cartesian mesh are eliminated. Additionally, the {\sc healpix} sphere is rotated in each timestep to wash out any remaining alignment effects. 

It is fully MPI-parallelized by employing asynchronous communication, i.e. as soon as a ray leaves the local domain it is sent to its neighbor. This has the advantage that domains without any point sources of their own are involved as soon as possible instead of having to wait until the neighbor completely sampled its domain. 

Ionising radiation is modeled in a photon conservative, mesh resolution independent fashion. Photon conservation is guaranteed by an incremental attenuation of the photon flux $\mathrm{d} N = N \ ( 1 - \mathrm{e}^{-\mathrm{d}\tau} )$, where the flux $N$ that enters a neighboring cell $i+1$ is $N_{i+1} = N_{i} - \mathrm{d}N_i$. $N$ is given as an absolute photon number $N = P \Delta t$, with $P$ in $\mathrm{s^{-1}}$. An ionization rate is calculated by exploiting the fact that $\mathrm{d}N$ is a dimensionless quantity which implicitly takes the dimensionality $m$ of the size of the cell $ V_{\mathrm{cell}} = d^m$ into account. The ionization rate is then given as $k_{ion} = \mathrm{d}N_{\mathrm{ion}} / ( n_{H} V_{\mathrm{cell}} \ \Delta t )$. The speed of the I-front expansion is captured by enforcing that the change in the abundance of atomic hydrogen, $\left| \mathrm{\Delta} x_{\mathrm{H},new} - \mathrm{\Delta} x_{\mathrm{H},old} \right| = \mathrm{\Delta x_{H}}$ is smaller than $10\%$. 

Additionally, {\sc fervent} is fully coupled to a reduced chemical network \cite{Glov12,Glov10,Glov07a,Glov07b,Walc14}. It explicitly tracks the species $\mathrm{CO}$, $\mathrm{C^+}$, $\mathrm{H^+}$, $\mathrm{H}$ and $\mathrm{H_2}$ which allows for a realistic chemical composition and hence a direct way to compare to observations. Moreover, it accounts for most heating processes by non-ionizing radiation, such as UV photodissociation, vibrational pumping and photoelectric heating.

\subsection{{\sc flash-treeray} (R. W\"unsch)}

{\sc treeray} (W\"unsch et al., {\it in prep.}) is a new radiation transport method combining a tree code with reverse ray-tracing. It has been implemented into the hydrodynamic code {\sc flash} \citep[][version 4.2.1 used in this work]{Fryx00}. The {\sc treeray} method has several advantages that make it suitable for being coupled with hydrodynamic codes. Particularly, (i) it is relatively fast (costs are similar to solving for self-gravity), (ii) the calculation time does not depend on number of sources (making it ideal for diffuse radiation treatment), (iii) ray-tracing has the highest resolution close to the point where the intensity is calculated, and (iv) it is relatively easy to parallelize this algorithm on distributed memory architectures. The disadvantage is that distant sources and regions of absorption are smoothed over a larger volume leading to the artificial numerical diffusion of radiation.

The algorithm kernel consists of three steps. In the first one, sources of radiation and the absorbing gas are mapped onto the octal tree and the emission and absorption coefficients are calculated for each tree node. In the second step, the tree is traversed for each grid cell and tree nodes are mapped onto rays going in all directions. The ray directions are obtained using the {\sc healpix} library for uniform tesselation of the surface of a sphere \citep{Gors05}, and the mapping coefficients are proportianal to the volume of the intersection of each tree node with the cone associated with the ray. In this way, no absorbing or emitting gas is omitted. In the third step, the one-dimensional radiation transport equation is solved along each ray. This computational kernel is called in each hydrodynamic time-step iteratively, until the radiation field converges, in order to account for regions irradiated from several different directions.

Various physical processes can be implemented into the code by providing prescriptions for the absorption and the emission coefficients. In this work, we use a simple on-the-spot approximation with only one source emitting a constant number of EUV photons per unit time and the number of absorbed photons given by the recombination rate to higher-than-ground levels.

\subsection{{\sc heracles} (P. Tremblin)}

{\sc heracles}\footnote{http://irfu.cea.fr/Projets/Site\_heracles} \citep{Gonz07,Audi11} is a 3D\footnote{ Here used as 1D only} hydrodynamical code used to simulate astrophysical fluid flows. It uses a finite volume method on fixed grids to solve the equations of hydrodynamics, MHD, radiative transfer and gravity. This software is developed at the Service d'Astrophysique, CEA/Saclay as part of the COAST project and is registered under the CeCILL license.

{\sc heracles} simulates astrophysical fluid flows using a grid based Eulerian finite volume Godunov method. It is capable of simulating pure hydrodynamical flows, magneto-hydrodynamic flows, radiation hydrodynamic flows (using either flux limited diffusion or the M1 moment method), self-gravitating flows using a Poisson solver or all of the above. {\sc heracles} uses cartesian, spherical and cylindrical grids. Current ongoing developments include a multi-grid method and a multi-group scheme for the radiative transfer.

The ionization scheme is described in \citet{Tremb11} and was applied to the study of the formation of pillars and globules at the interface of H~{\sc ii} regions and turbulent molecular clouds \citep[see][]{Tremb12a,Tremb12b}.

\subsection{{\sc pion} (J. Mackey)}
\textsc{pion} is an Eulerian magnetohydrodynamics code that solves either the Euler equations of gas dynamics \citep{Mack10} or the ideal magnetohydrodynamics equations \citep{Mack11} on a uniform fixed mesh in Cartesian (1D, 2D, 3D), cylindrical (2D axisymmetric), or spherical (1D) coordinates. It uses a finite volume, shock-capturing integration scheme \citep{Fall98} with geometric source terms to account for curvilinear coordinates when needed \citep{Fall91}. Abundances of chemical species and tracers are passively advected with the flow. \textsc{pion} is written in \texttt{C++}, is designed to be as modular as possible, is parallelised with MPI by dividing the domain into $N$ subdomains ($N$ must be a power of 2), each controlled by one MPI process, and scales well to at least 1024 cores for 3D simulations.

Radiative transfer of ionizing and non-ionizing photons from point sources and sources at infinity is followed using a short characteristics raytracer to calculate column densities. This is coupled to a microphysics module to solve the rate equations for chemical species and their associated heating or cooling. Scattered and re-emitted photons are treated using the ``on-the-spot'' approximation, i.e.\ they are assumed to be re-absorbed locally. A photon-conserving formulation of the ionization rates is used and spectral hardening of ionizing radiation with optical depth is included \citep[following][]{Mell06b}. The non-equilibrium ionization of H is integrated together with the thermal evolution of the gas using \textsc{cvode} \citep{Cohe96}, a high-order integrator for coupled differential equations, set to use backward differencing with Newton iteration. The ionization and heating/cooling source terms are integrated explicitly in the finite volume scheme, using an innovative algorithm that preserves the second order time-accuracy of the numerical scheme and dramatically reduces the computation required for a given error tolerance \citep{Mack12}.

\subsection{{\sc ramses-lampray} (T. Haugb\o lle, T. Frostholm)}
The radiative transfer module {\sc lampray} (Long characteristics AMR Parallel RAY tracing) is implemented into a derivative of the {\sc ramses} cosmological code \citep{Teys02,From06}, which has been adapted to the detailed study of star formation. The code solves the MHD equations on a fully threaded tree (FTT) with support for self-gravity. Important additional physics modules compared to the original {\sc ramses} code include accreting sink particles, coupling to the astrochemistry framework KROME, and many smaller changes to improve the stability and quality of the HLLD MHD solver in the case of supersonic flows. For the D-type test the hydrodynamics is solved using a second order MUSCL scheme with an HLLD solver and an isotropic monotonized central slope limiter. 

The radiative transfer module employs ray tracing directly on the adaptive mesh. These are long characteristics rays covering the entire computational domain. The technique is made computationally feasible by, in addition to the FTT, also using a separate ray-oriented domain decomposition, in which the radiative transfer problem can be solved efficiently in parallel. A photon-conserving second-order accurate TSC interpolation between the two domains is used to calculate densities and abundances along the ray, and deposit the ionization and heating rates. The ionization solver is based on the {\sc c$^{2}$-ray} method \citet{Mell06b}. In a given timestep first the ray geometry is established and the MPI and sorting keys are set up linking effectively cell centers with points along the ray. Then the rates in the timestep are found using an iterative method, that repeatedly computes the average ionization rate and the corresponding changes in abundance, until the two converge. The result is that in principle arbitrarily large time steps can be taken, and the method is only limited by the Courant condition imposed by the fluid dynamics. 

The ray scheme allows for a completely arbitrary placement of rays, and only domains that actually intersect with rays store knowledge about them, making the method scalable to thousands of cores. The flexible placement of rays in addition makes it possible to use a mixture of rays and solve simultaneously for the diffuse and point-source components of the radiation field. 

\subsection{{\sc ramses-rt} (S. Geen \& J. Rosdahl)}

{\sc ramses-rt}\footnote{The {\sc ramses-rt} implementation is publicly available, as a part of the {\sc ramses} code (https://bitbucket.org/rteyssie/ramses)} \citep{Rosd13} is a radiation-hydrodynamical extension of the {\sc ramses} cosmological code. {\sc ramses} \citep{Teys02} solves the equations of gravitational-hydrodynamics with a second-order unsplit Godunov solver  on an adaptive mesh, using a fully threaded tree structure. {In this paper, the hydro-solver specified by \citet{Toro94} has been adopted.} The code is fully (MPI) parallel.

In {\sc ramses-rt}, RT is integrated into the structural framework of {\sc ramses} and coupled to the hydrodynamics via interactions with hydrogen and helium. The RT is solved on the AMR grid with a first-order moment method using the M1 closure for the Eddington tensor. This strategy has the advantages that the computational load is invariant with the number of radiation sources (moment method) and the radiation transport solver is local in space (M1 closure), i.e. it requires no information outside the local volume when advecting the photon fluid between cells.

The radiation-thermochemistry of hydrogen and helium is solved in {\sc ramses-rt} assuming non-equilibrium, where the ionised fractions of those species are tracked explicitly as passive scalars that are advected with the gas flow.

For adaptability of the code, and because the implementation is designed in large part to simulate galactic and extragalactic scales, where radiation sources are highly dynamic, the radiation is advected with an explicit solver, which is subject to a Courant condition for the time step. The solution to overcoming the extremely small implied time-step is to use the so-called reduced speed of light approximation \citep{Gned01}: the speed of light is simply slowed down by some factor, typically two or three orders of magnitude, such that the hydrodynamical time-step length is roughly maintained. In the {\sc ramses-rt} tests described here, we reduce the speed of light by a factor $10^{-2}$ compared to the real value.

\subsection{{\sc sedna} (R. Kuiper)}
The ionization module named {\sc sedna} is currently under development by Rolf Kuiper. The module is coupled to the static grid version of the open source MHD code {\sc pluto} \citep{Mign07}. Up to now, the module works with static, rectangular, uniform and non-uniform grids in 1D-3D Cartesian, cylindrical, and spherical coordinates.

Similar to the hybrid radiation transport solver, introduced in \citet{Kuip10b} and \citet{Kuip13}, the total ionizing radiation field is split into a direct ionization component and a diffuse radiation field. The ionization by direct irradiation from either a point source at the center of the spherical domain or a plane parallel flux in Cartesian geometry is computed by a ray-tracing step along the first coordinate axis. For the secondary radiation field, the module can either make use of the so-called ``on-the-spot'' approximation, or the ionization of EUV photons created by direct recombination into the ground state can be computed using the flux-limited diffusion approximation. Ongoing development of the module is based on the numerical code descriptions by \cite{York96} and \cite{Rich97,Rich00} as well as the lecture course material in \citet{Kudr88}. The ionization module {\sc sedna} can be coupled to the radiation transport solver mentioned above to allow the determination of gas and dust temperature in dusty ionized and non-ionized regions, respectively. Future development might include dust scattering and FUV radiation fields, which contribute to the photo-heating of gas as well as to Carbon ionization.

The main purpose of the ionization module {\sc sedna} will be its application in modeling the formation and feedback of massive (proto-)stars. In this sense, the module comprises an extension to previous numerical studies regarding radiation pressure \citep{Kuip10a,Kuip11,Kuip12,Kuip13a}, stellar evolution \citep{Kuip13b}, and protostellar outflow feedback \citep{Kuip15}.

\subsection{{\sc seren} (D. Hubber, T.G. Bisbas)}
{\sc seren} \citep{Hubb11} is an SPH code designed for star formation, planet formation and star cluster problems.  {\sc seren} uses a conservative `grad-h' SPH implementation to model the hydrodynamics, a Barnes-Hut tree to model self-gravity, sink particles to model accretion onto protostars and several algorithms to model protostellar and stellar feedback \citep[i.e.][]{Stam12}.

\citet{Bisb09} used the {\sc healpix} algorithm \citep{Gors05} to tessellate all surrounding directions into discrete vectors of equal solid angle that cover the whole surface.  In that method, ionising radiation is propagated along each {\sc healpix} ray, calculating the position of the ionisation front and assuming ionisation equilibrium and neglecting the diffuse field. The trapezium method is used to calculate the density at various `evaluation points' along the ray following the radiation.  The distance to the next evaluation point is given by $\Delta r = f_1\,h$ where $f_1$ is a dimensionless constant of order unity and $h$ is the smoothing length calculated at the previous evaluation point.  At the final evaluation point, a bisection method is used to accurately determine the location of the final ionisation front.  The temperature of particles is smoothed around the ionisation front to prevent the two fluids (hot and cold) from becoming separated and forming a gap \citep[e.g. see][]{Pric08}.  The angular resolution of the ionising radiation can be refined at any point by splitting a ray into four child rays, which then compute the rest of the ionisation integral independently.  A ray is split when the width of the ray cone exceeds some fraction of the local smoothing length.  This matches the ray resolution to the gas particle resolution.

{\sc seren} now contains an updated version of this algorithm with two main improvements and optimisations.
\begin{enumerate}
\item When creating the {\sc healpix} tessellation, the particles are divided into linked lists along each ray.  When the rays are split to improve the resolution, the linked lists are also split along each ray amongst the four child rays.  When the radiation is propagated along each ray, the step-size between evaluation points is taken as the minimum of the smoothing lengths between the previous and next particles.  This ensures that the next evaluation point does not `over-shoot' if there is suddenly a dense region such as a high density clump or a shock.  This means the bisection iteration to find the ionisation front is now performed more accurately with fewer steps.
\item When walking along the linked lists, if the previous and next particles have very similar properties to within some tolerance, then the density can be extrapolated from the particles rather than performing another (expensive) tree-walk.  This speeds up the calculation, depending on the chosen tolerance.
\end{enumerate}

Both of these optimisations lead to faster run times and a more accurate determination of the location of the ionisation front.  For roughly uniform density distributions, both methods give the same results.  For strong density contrasts, particularly near the ionisation front (which can be common later in the simulation), the results may diverge due to the different accuracies of finding the ionisation front position.  We perform all ionisation tests using the latter implementation only.

\subsection{{\sc torus} (T. Haworth)}
\label{amr}
\textsc{torus} is principally an Adaptive Mesh Refinement (AMR) Monte Carlo radiation transport code capable of continuum, atomic line, non-LTE molecular line radiation transport  \citep[e.g.][]{Harr00, Kuro06, Rund10} and most recently coupled radiation transport and photodissociation region chemistry (Bisbas et al. 2015, {\it submitted}).  It was developed to treat radiation hydrodynamics problems using operator split photoionization and hydrodynamics by \citet{Hawo12} as follows.

\textsc{torus} uses a flux conserving, finite difference, total variation diminishing hydrodynamics scheme. It uses the Superbee flux limiter \citep{Roe85} and includes a Rhie--Chow interpolation scheme to prevent odd-even decoupling \citep{Rhie83}. 

Monte Carlo photoionization solves for the time-averaged energy density $dU$ in each cell by propagating constant-energy $\epsilon$ packets of photons over the computational grid and counting their path lengths $l$ through each cell of volume $V$ 
\begin{equation}
	dU = \frac{4\pi J_{\nu}}{c} d \nu = \frac{\epsilon}{c\Delta t}
        \frac{1}{V} \sum_{d\nu} l,
\label{energydensityMC}
\end{equation}
where $\epsilon$  is the total source luminosity divided by the number of photon packets used. 

The estimated energy densities are used in the photoionization equilibrium equation \citep{Oste89} to solve for the ionization balance, which in terms of Monte Carlo estimators is 
\begin{equation}
	\frac{n(X^{i+1})}{n(X^i)} = \frac{ \epsilon}{\Delta t V \alpha(X^i) n_{\rm{e}}} \sum\frac{l \bar{\sigma}_{\nu}(X^i)}{  h\nu},
	\label{ionBalanceMC}
\end{equation}
where $n(X^i)$, $\alpha(X^i)$, $\bar{\sigma}_{\nu}(X^i)$ and $n_{\rm{e}}$ are the number density, recombination coefficient and absorption cross section of ion $X^i$ and the electron density respectively. The ionization fraction yields the temperature and therefore pressure under the simple thermal prescription used in this paper. To remain consistent with the other codes the OTS approximation is employed by terminating the propagation of a photon packet after its first absorption and by using the case B recombination coefficient. Using operator splitting, hydrodynamics and photoionization calculations are performed sequentially until the simulation end time.  The advantage of this approach is that many complex features usually only treated by dedicated radiation transport codes can be included in RHD applications \citep[e.g. the diffuse radiation field,][]{Hawo12}. The disadvantage is that it is computationally expensive, but this is overcome owing to the efficient parallelization of Monte Carlo radiation transport \citep{Harr15}.

\section{The D-type Benchmark Test}%
\label{sec:test}

\subsection{Initial Physical conditions}
\label{ssec:ic}

For the purposes of the test we will use a simplified isothermal equation of state for both the ionized and the neutral medium. The treatment of the interface where the two media meet is left to each numerical method. The test is purely hydrodynamical i.e. it includes no gravity or magnetic fields. Due to the nature of Eqn. (\ref{eqn.raga}) we will run two tests: i) to examine the early phase of the D-type expansion i.e. where the Spitzer approximation (Eqn.~\ref{eqn.Spitzer}) is applicable and ii) to examine the later phase of the D-type expansion, i.e. where the Raga expression (Eqn.~\ref{eqn.raga}) is applicable. 

In both early and late phase tests, we consider a uniform region containing pure hydrogen and an ionizing source emitting $\dot{\cal N}_{_{\rm LyC}}=10^{49}$ photons per second placed at the origin. The position of the source defines the origin of a Cartesian coordinate system. The density of the gas (hydrogen only) is taken to be $\rho_{\rm o}=5.21\times10^{-21}\,{\rm g}\,{\rm cm}^{-3}$.  The temperature and the sound speed of the ionized gas are taken to be $T_{\rm i}=10^4\,{\rm K}$ and $c_{\rm i}=12.85\,{\rm km}\,{\rm s}^{-1}$ respectively.

\subsubsection{Early phase}

For the early phase expansion test the temperature of the neutral gas is initialized to $T_{\rm o}=10^2\,{\rm K}$ corresponding to a sound speed of $c_{\rm o}=0.91\,{\rm km}\,{\rm s}^{-1}$. For the SPH calculations, we assume the dense gas takes the form of a cloud of finite radius, with total mass of $M_{\rm cl}=640\,{\rm M}_{\odot}$. The radius of the cloud is therefore $R_{\rm cl}=1.257\,{\rm pc}$, or $R_{\rm cl}=4R_{\rm St}$. We evolve the simulation until $t_{\rm end}=0.141\,{\rm Myr}$. At this time the ionization front has reached the edge of the domain (or the cloud in the SPH runs). The upper panel of Fig.~\ref{fig:analytical} plots the analytical equations described in Section \ref{sec:theory} for the early phase expansion.

\subsubsection{Late phase}

Equation \ref{eqn.rstag} gives the stagnation radius as a function of the initial Str{\"o}mgren radius. 
\begin{eqnarray}
R_{_{\rm STAG}}=\left(\frac{c_{\rm i}}{c_{\rm o}}\right)^{4/3}R_{\rm St}\,,
\end{eqnarray}
As we discuss in \S\ref{sssec:sphsetup}, the number of SPH particles required for the late phase test may lead to prohibitively high computational expenses (see Eqn.~\ref{eqn:sphparticles} in particular where there is a dependency on $\left(\frac{c_{\rm i}}{c_{\rm o}}\right)^{4}$).
We will therefore adopt a `neutral' temperature of $T_{\rm o}=10^3\,{\rm K}$ while keeping $\mu_{\rm o}=1$. The corresponding sound speed is then $c_{\rm o}=2.87\,{\rm km}\,{\rm s}^{-1}$.  As for the early-phase calculations, the initial Str{\"o}mgren radius is $R_{\rm St}=0.314\,{\rm pc}$. The stagnation radius $R_{_{\rm STAG}}=2.31\,{\rm pc}$ is obtained at $t_{_{\rm STAG}}\simeq3.0\,{\rm Myr}$ which defines the time we terminate the simulation. 

\begin{figure}
\centering
\includegraphics[width=0.45\textwidth]{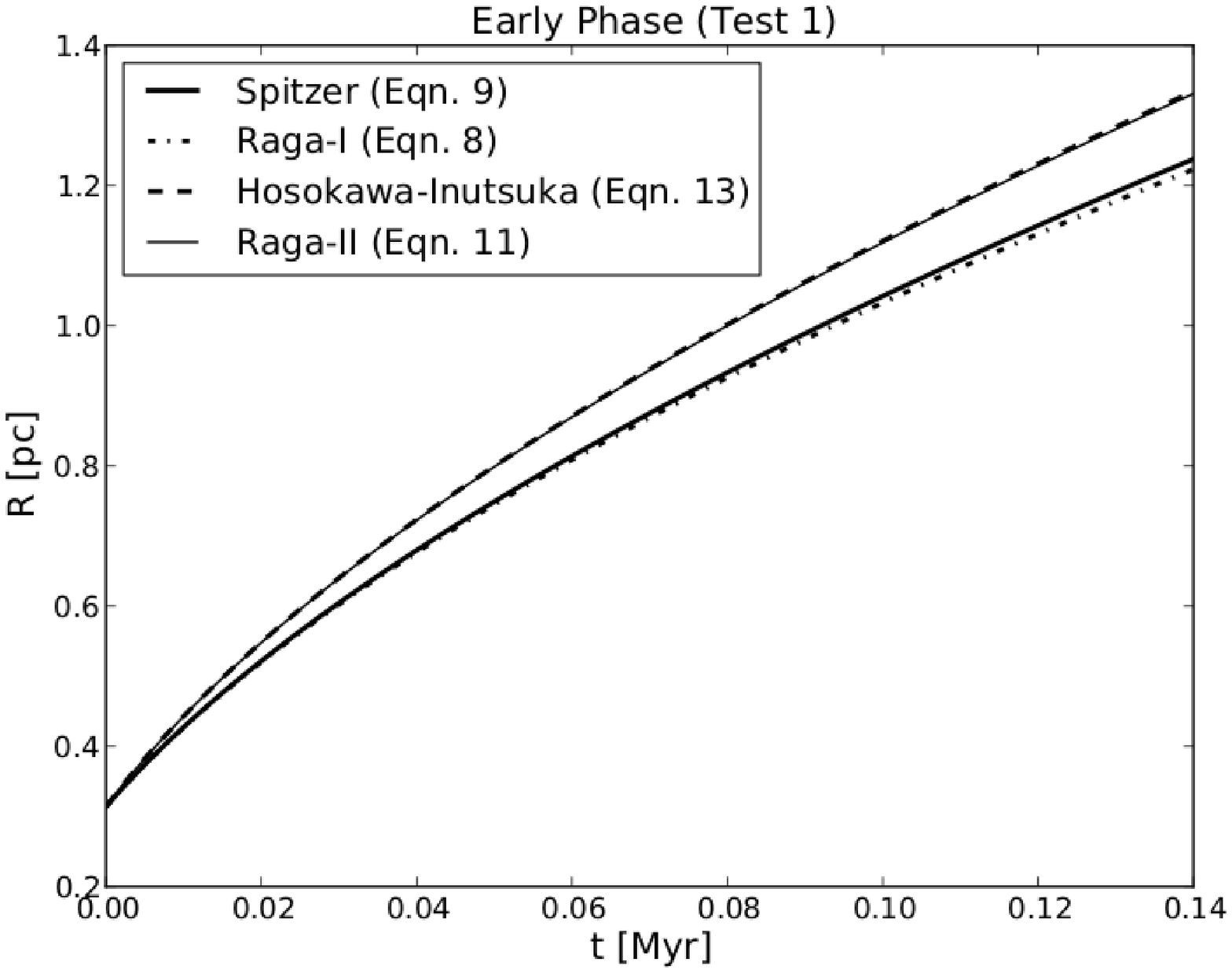}
\includegraphics[width=0.45\textwidth]{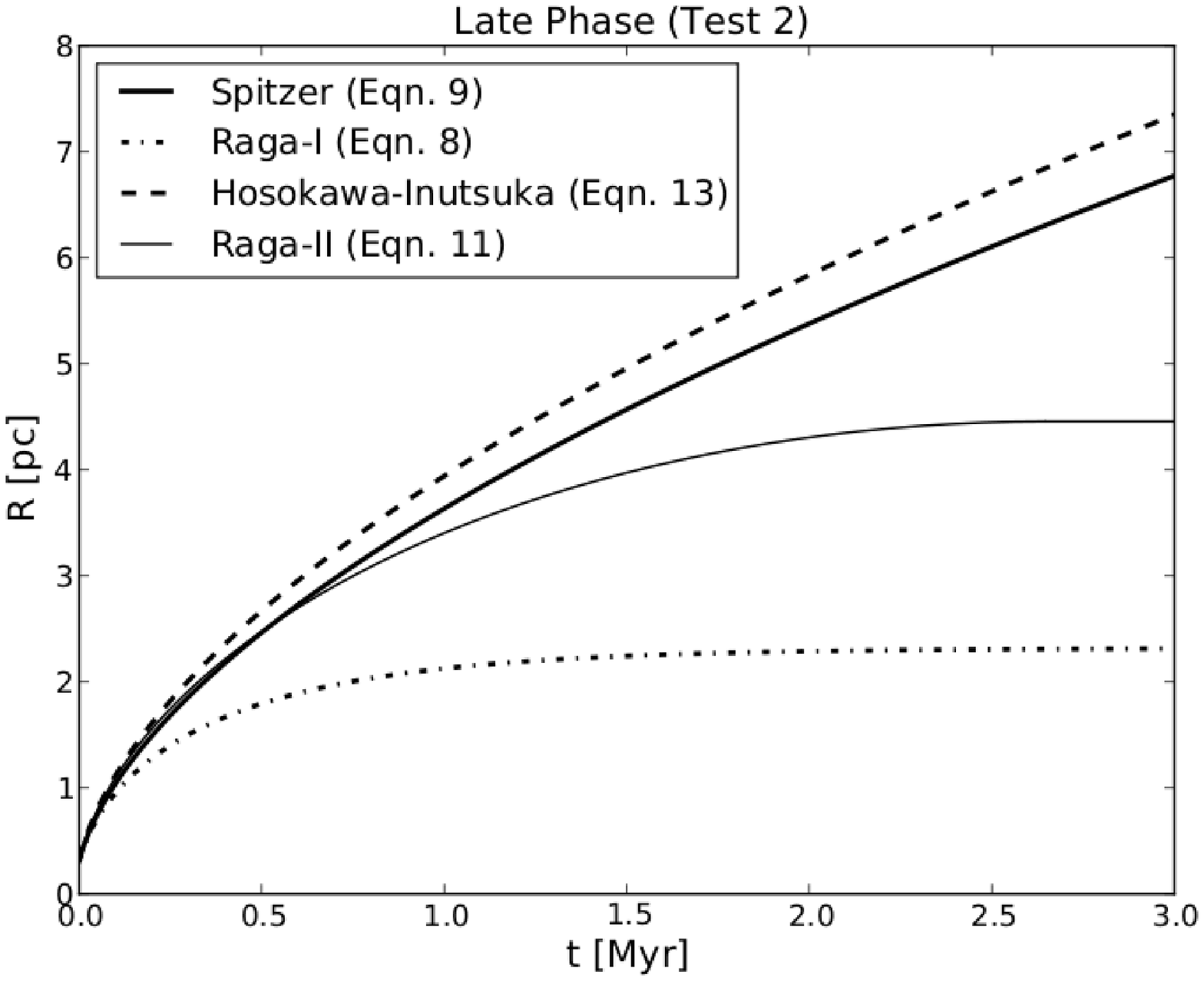}
\caption{ Analytical expressions of Eqns.~\ref{eqn.raga} (Raga-I), \ref{eqn.raga2diff} (Raga-II), \ref{eqn.Spitzer} (Spitzer), and \ref{eqn:HI} (Hosokawa-Inutsuka) for the `early phase' (upper panel) and for the `late phase' tests. The $x$-axis is the time (Myr) and the $y$-axis the position of fronts (pc).}
\label{fig:analytical}
\end{figure}

For the late phase SPH calculations, the cloud has mass $M_{\rm cl}=8\times10^3\,{\rm M}_{\odot}$ and $R_{\rm cl}=2.91\,{\rm pc}$. The lower panel of Figure \ref{fig:analytical} plots the analytical equations described in Section \ref{sec:theory} for the late phase expansion.

\subsection{Numerical setup and configuration}
\label{ssec:setup}
\subsubsection{SPH setup}
\label{sssec:sphsetup}

We choose a slightly bigger cloud radius than the stagnation radius to avoid the effect of an expanding shell in vacuum when $t=t_{_{\rm STAG}}$. Let $g$ be this additional factor determining the extra size of the cloud radius, 
\begin{eqnarray}
R_{\rm cl}=g R_{_{\rm STAG}}
\end{eqnarray}
and let $f\equiv g\left(\frac{c_{\rm i}}{c_{\rm o}}\right)^{4/3}$.

Suppose that $R_{\rm cl}=fR_{_{\rm St}}$, where $R_{\rm cl}$ is the radius of the spherical cloud, $R_{_{\rm St}}$ is the Str{\"o}mgren radius and $f>1$ is a user-defined factor. $f$ describes the size of the Str{\"o}mgren sphere in comparison with the size of the spherical cloud. 

The density at $t=0$ is homogeneous throughout the cloud. Therefore,
\begin{eqnarray}
\rho_{\rm o}=\frac{f^3 \dot{\cal N}_{_{\rm LyC}} m^2}{M_{\rm cl} \alpha_{_{\rm B}}}.
\end{eqnarray}

Let $N_{_{\rm SPH}}^{\rm St}$ be the number of SPH particles consisting the Str{\"o}mgren sphere. All SPH particles have the same mass, $m_{_{\rm SPH}}$. Then

\begin{eqnarray}
\rho_{\rm o}=\frac{3 M_{_{\rm St}}}{4 \pi R_{_{\rm St}}^3}=\frac{3 N_{_{\rm SPH}}^{\rm St} m_{_{\rm SPH}}}{4 \pi R_{_{\rm St}}^3}
\end{eqnarray}
and
\begin{eqnarray}
\rho_{\rm o}=\frac{3 M_{\rm cl}}{4 \pi R_{\rm cl}^3}=\frac{3 N_{_{\rm SPH}}^{\rm cl}m_{_{\rm SPH}}}{4 \pi R_{\rm cl}^3}.
\end{eqnarray}
Combining the above, we obtain $N_{_{\rm SPH}}^{\rm cl}=f^3 N_{_{\rm SPH}}^{\rm St}$. This equation gives the total number of SPH particles for a specific size and resolution of the Str{\"o}mgren sphere. In all cases we will use a pre-settled particle distribution i.e. a `glass' structure, which minimizes (but does not eliminate) the numerical noise.

For the test examining the early phase we will use $f=4$ and $N_{_{\rm SPH}}^{\rm St}=10^4$. Therefore the total number of SPH particles is taken to be $N_{_{\rm SPH}}^{\rm cl}=6.4\times10^5$ particles. We also use $m_{_{\rm SPH}}=10^{-3}\,{\rm M}_{\odot}$, which therefore implies $M_{\rm cl}=640\,{\rm M}_{\odot}$. 

For the test examining the late phase, the total number of SPH particles is:
\begin{eqnarray}
\label{eqn:sphparticles}
N_{_{\rm SPH}}^{\rm cl}=g^3\left(\frac{c_{\rm i}}{c_{\rm o}}\right)^4 N_{_{\rm SPH}}^{\rm St}\,.
\end{eqnarray}
Using $g=1.26$ (because at this value $g^3\simeq2$) and $N_{_{\rm SPH}}^{\rm St}=10^4$ we obtain a total of $N_{_{\rm SPH}}^{\rm cl}=8\times10^6$ SPH particles, each one having mass $m_{_{\rm SPH}}=10^{-3}\,{\rm M}_{\odot}$.

\subsubsection{3D grid setup}%

The radiation source is placed at the origin, as before. The three-dimensional Eulerian calculations are run for a single octant with a spatial resolution of $128^3$ grid zones. Reflecting boundary conditions are used in the negative directions, and zero gradient boundary conditions in the positive directions. An octant runs from $\{x,\,y,\,z\}\in[0,3.874\times10^{18}]$\,cm, corresponding to $4R_{\mathrm{St}}$ ($R_{\mathrm{St}}=0.9685\times10^{18}\,\mathrm{cm}=0.314\,$ pc).

For the late phase test the simulation was set up in a similar way, but with the simulation domain now extending to $1.26\, R_\textsc{stag} = 2.91$\,pc to agree with the SPH setup (Sec.~\ref{sssec:sphsetup}). The ISM density is the same, but the neutral gas temperature is initialized to $10^3\,{\rm K}$ instead of $10^2\,{\rm K}$, so the pressure ratio between ionised and neutral gas is reduced to 20 in the initial conditions.

\subsubsection{1D simulations}
\label{sssec:1dgrid}
For 1D simulations, contributors were encouraged to use any algorithm available for solving spherically symmetric problems. The simulation domain was typically set to a somewhat larger value than $4R_\mathrm{St}$ so that the shock front remained inside the grid until the finishing time of the simulation. For the early phase test, a maximum radius of 1.5 pc was sufficient, while for the late phase test the maximum radius was taken to be 5 pc. Numerical schemes used include a uniform grid ({\sc pion},{\sc sedna}, {\sc heracles}, {\sc torus}), adaptive Eulerian mesh ({\sc aquiline}), and a Lagrangian mesh ({\sc glide}). The different schemes resulted in a significant range of spatial resolutions from code to code.

\section{Results}
\label{sec:results}%

\subsection{1D runs}
\label{ssec:r1d}
\subsubsection{Early phase test (Test 1-1D)}
\label{sssec:1Dearly}

\begin{figure*}
\centering
\includegraphics[width=0.45\textwidth]{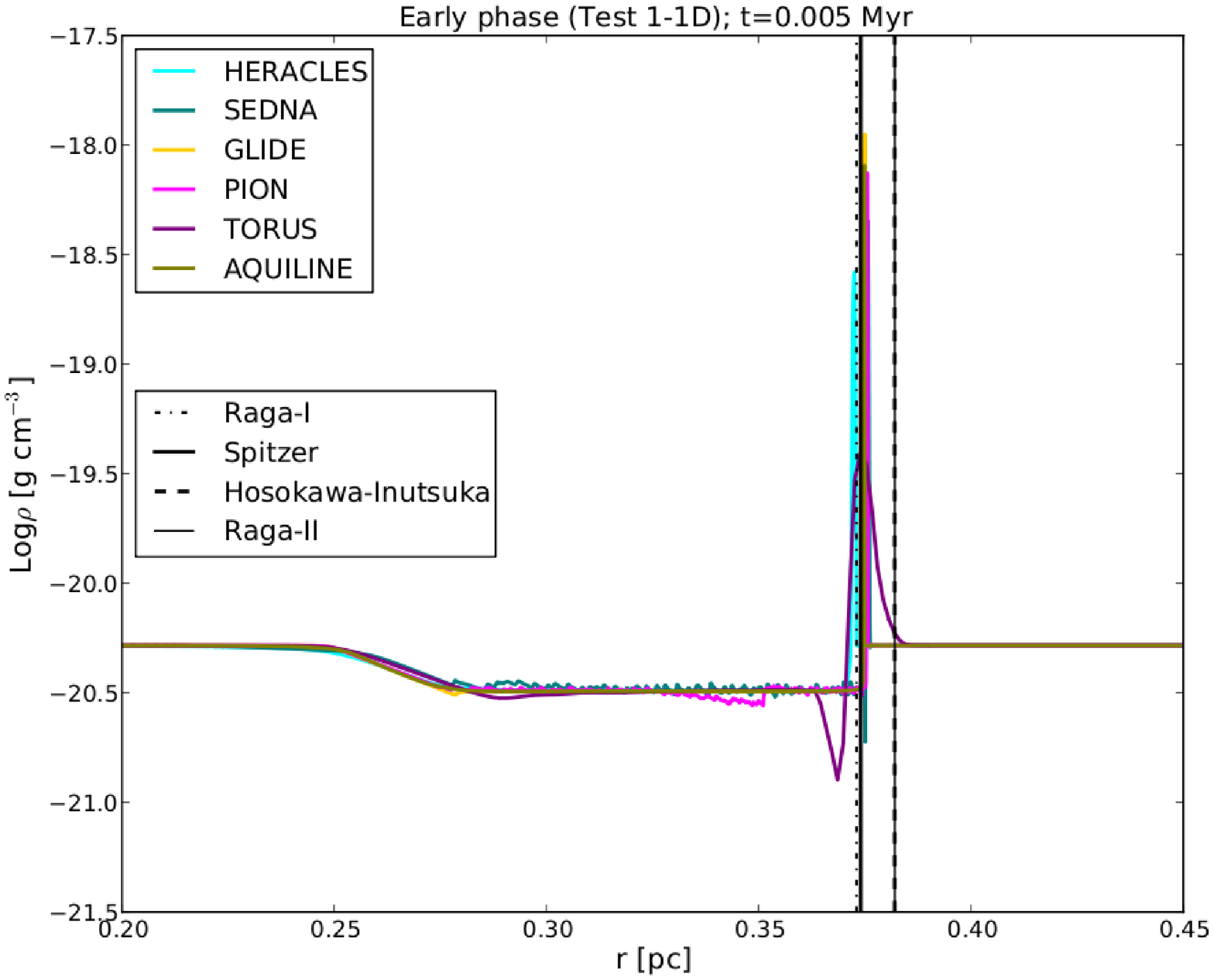}
\includegraphics[width=0.45\textwidth]{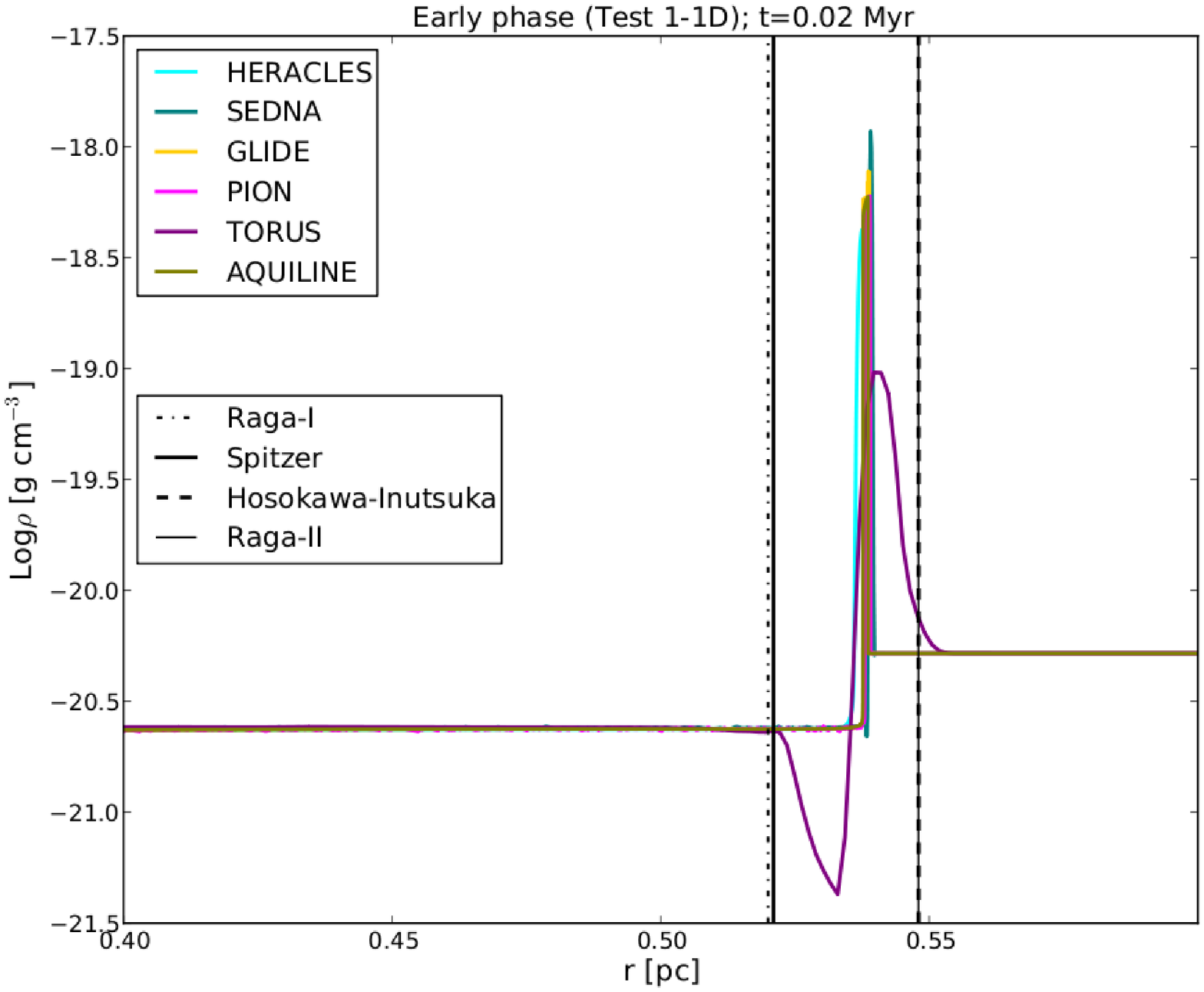}
\includegraphics[width=0.45\textwidth]{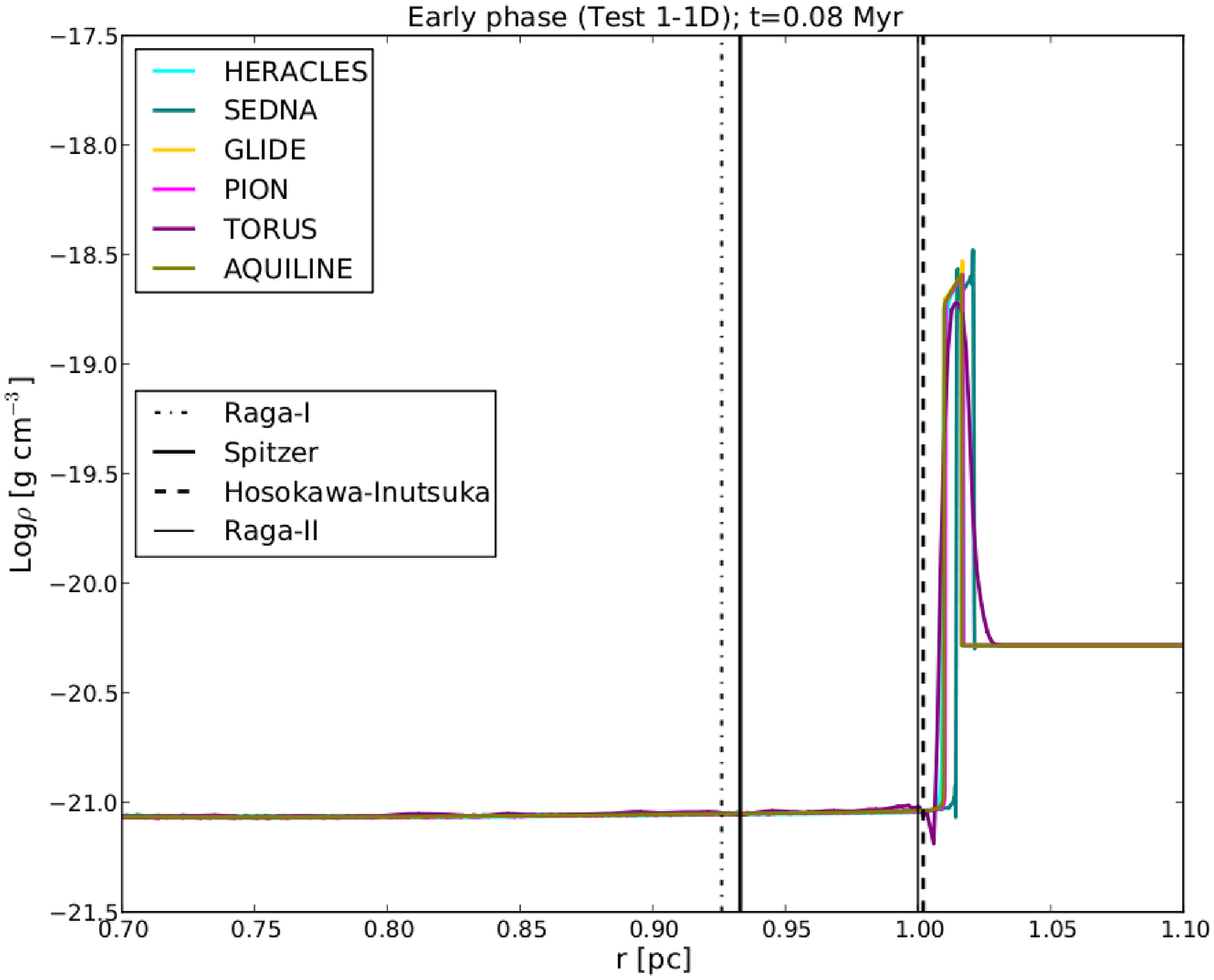}
\includegraphics[width=0.45\textwidth]{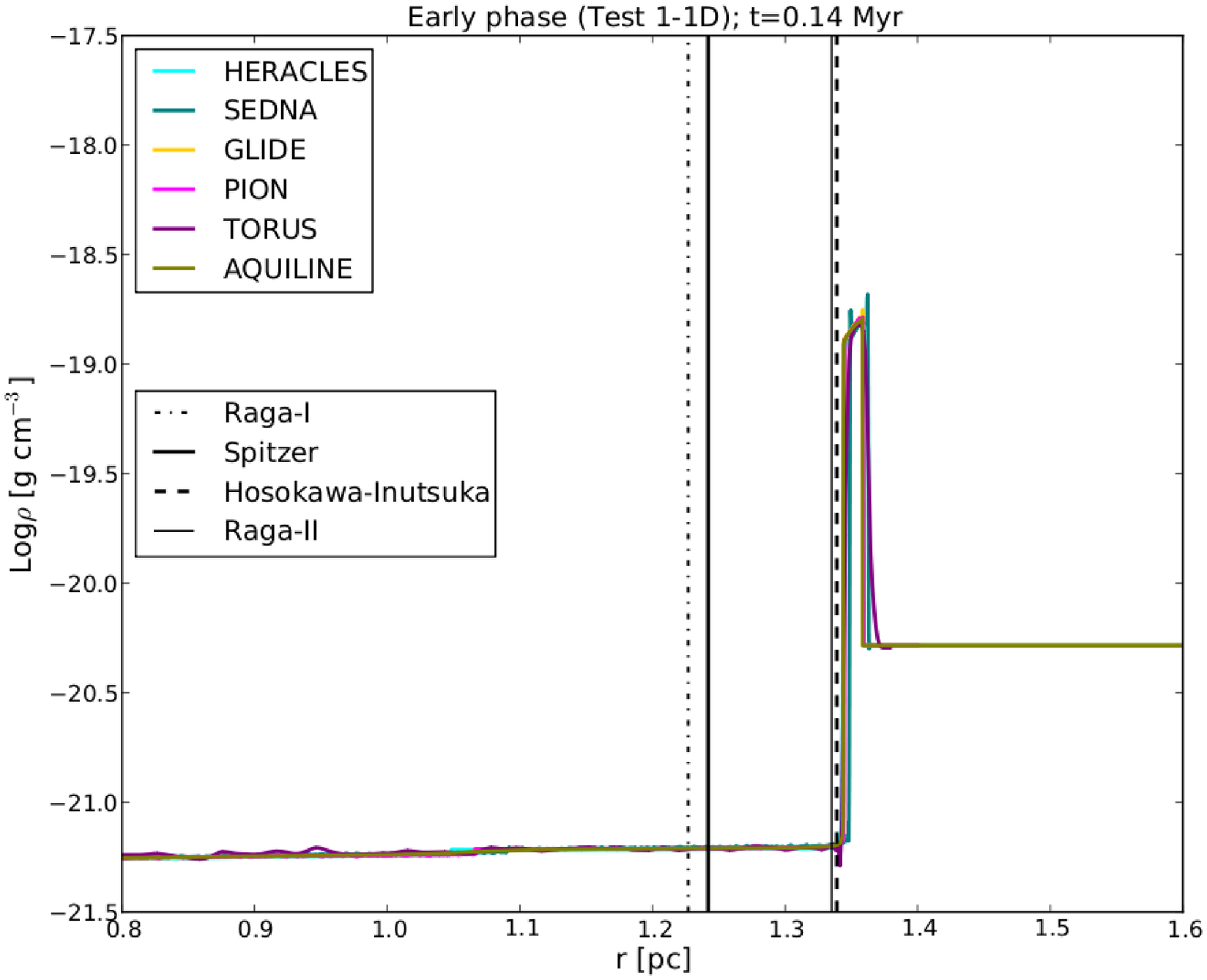}
\caption{ Density plots of the 1D codes for the early phase test. The panels correspond to different evolutionary times. The $x$-axis in these panels is the distance from the ionizing source (pc) and the $y$-axis is the density (${\rm g}\,{\rm cm}^{-3}$). We plot snapshots at $t=0.005\,{\rm Myr}$ (top left), $t=0.02\,{\rm Myr}$ (top right), $t=0.08\,{\rm Myr}$ (middle left), and $t=0.14\,{\rm Myr}$ (middle right). We find very good agreement between all codes. We see that at $t=0.005\,{\rm Myr}$ the codes reproduce the position predicted by the Spitzer equation (Eqn.~\ref{eqn.Spitzer}), whereas at the end of the simulation, $t=0.14\,{\rm Myr}$, the maximum density is located further than the Hosokawa-Inutsuka equation (Eqn.~\ref{eqn:HI}). Further details are discussed in \S\ref{sssec:1Dearly}}
\label{fig:1Dearly}
\end{figure*}

The six 1D hydrodynamical codes use a variety of grids specified in Table \ref{tab:allcodes}. In these runs, a variety of resolutions have been studied.  Where practicable, runs were performed at increasing resolution until the results converged and the resolution requirements turned out to be different for different codes.  In this section, we show only the highest resolution results, in order to provide a validated reference solution to which other results may be compared (see Table~\ref{tab:resol} for the resolution each 1D code used).

\begin{table}
\caption{ Number of cells used by the 1D participating codes for both early and late phase tests (Test 1-1D and Test 2-1D respectively).}
\centering
\label{tab:resol}
\begin{tabular}{l c}
\hline
\hline
Code name 		& No. grid cells \\
\hline
{\sc aquiline} 		&  32768 \\
{\sc glide}		&  378 \\ 
{\sc heracles} 		&  16834 \\ 
{\sc pion}		&  131072 \\ 
{\sc sedna}		&  16834 \\ 
{\sc torus}		&  1024 \\ 
\hline
\end{tabular}
\end{table}

Figure \ref{fig:1Dearly} shows the density as a function radius for the six 1D codes participating in the Test 1-1D. In this figure we plot four different evolutionary times, i.e. $t=0.005$, $t=0.02$, $t=0.08$ and $t=0.14\,{\rm Myr}$. We also indicate with vertical lines the position of the extent of the H~{\sc ii} regions as predicted by the various approximations presented in Section \ref{sec:theory}. 
All codes agree very well with each other; the peak density and shell width vary between codes because of the different spatial resolutions used. At early times the position of the thin shell agrees with the Spitzer solution (Eqn.~\ref{eqn.Spitzer}) but then moves towards the Hosokawa-Inutsuka (Eqn.~\ref{eqn:HI}) solution even overtaking it at late times. At $t=0.08$ and 0.14 Myr it is clear that Eqn.~\ref{eqn.Spitzer} no longer correctly describes the evolution. At this stage all codes at least partially resolve the shell width.

\begin{figure}
\centering
\includegraphics[width=0.45\textwidth]{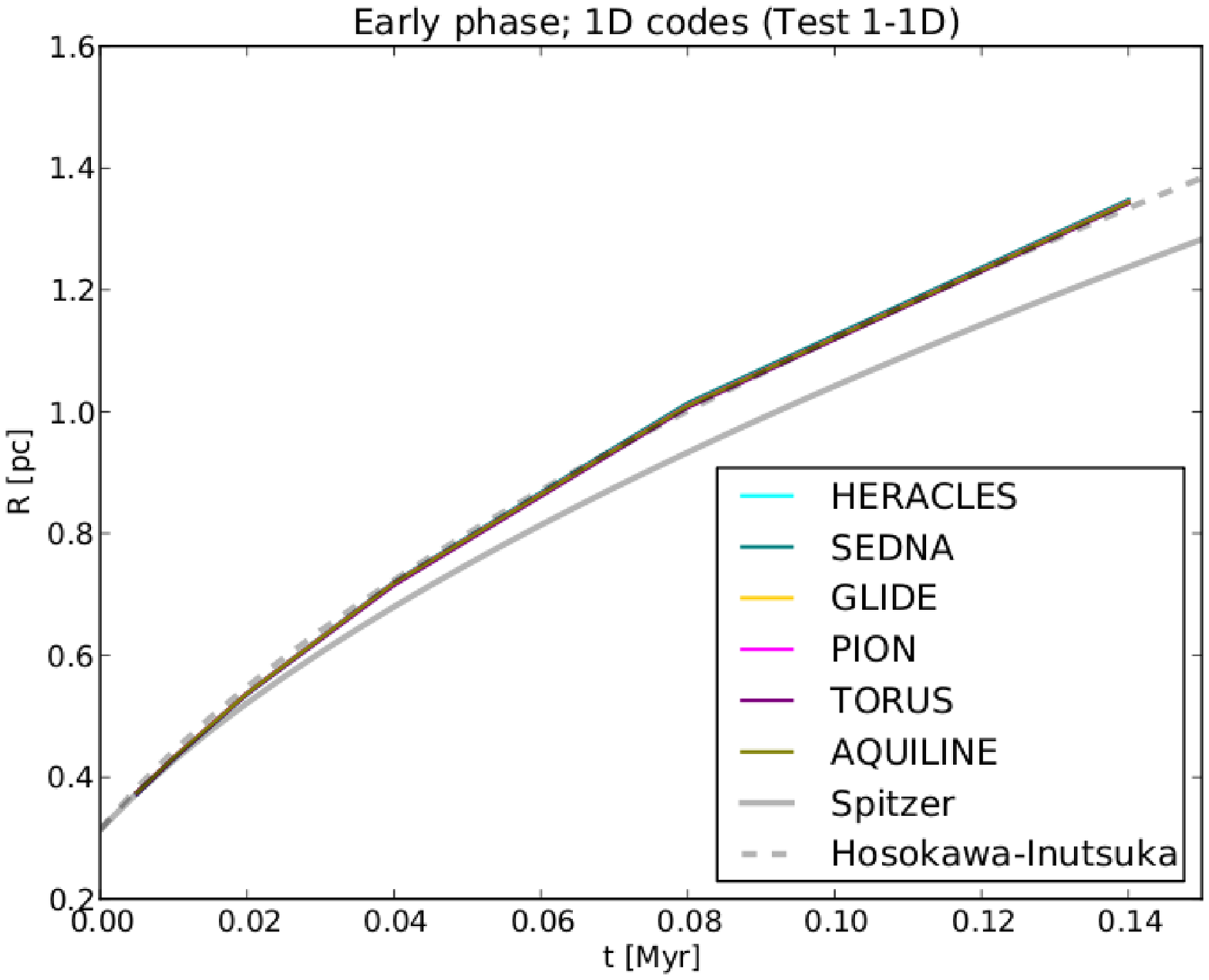}
\includegraphics[width=0.45\textwidth]{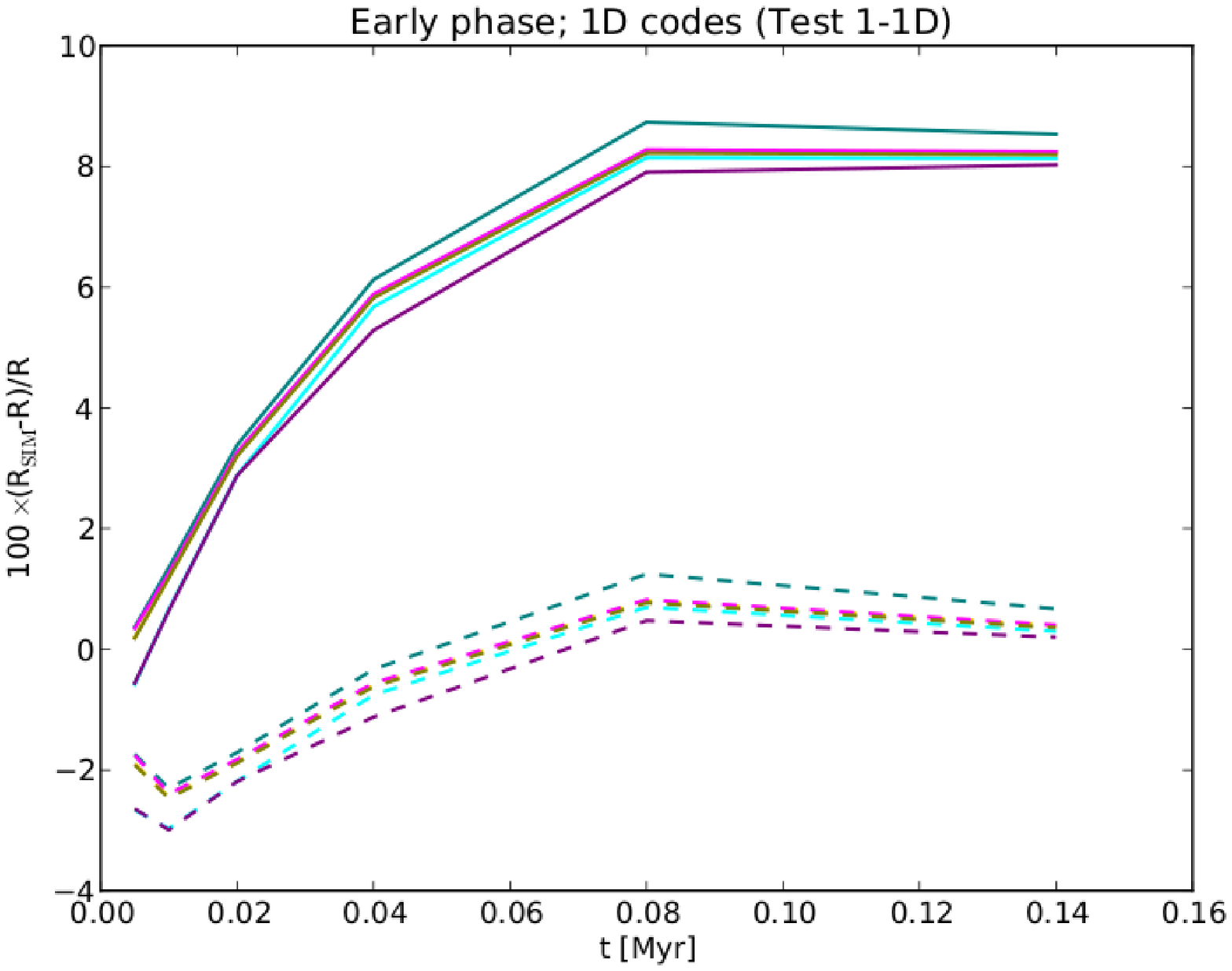}
\caption{ Position at which the ionization fraction is $x_{\rm i}=0.5$ as calculated by the 1D participating codes. The upper panel compares the codes with the Spitzer (Eqn.~\ref{eqn.Spitzer}; solid gray line) and the Hosokawa-Inutsuka (Eqn.~\ref{eqn:HI}; dashed gray line) equations. The lower panel shows the realtive error found between the numerical solution and either of the two analytical equations (solid lines correspond to the errors due to the comparison with the Spitzer equation, while dashed lines due to the Hosokawa-Inutsuka equation).}
\label{fig:1Dearlyerrors}
\end{figure}

The two panels in Fig.~\ref{fig:1Dearlyerrors} show the position of the the ionization front, defined as the position at which $x_{\rm i}=0.5$, compared to the Spitzer (Eqn.~\ref{eqn.Spitzer}) and the Hosokawa-Inutsuka (Eqn.~\ref{eqn:HI}) approximations. The lower panel shows the relative error of each code as compared to those two analytical approximations. This relative error is $\sim8\%$ for the Spitzer and $\lesssim1\%$ for the Hosokawa-Inutsuka approximation at $t\gtrsim0.08\,{\rm Myr}$, and so we conclude that all participating codes agree with the Hosokawa-Inutsuka approximation and none with the Spitzer formulation at this level of accuracy. 

In Table \ref{tab:early} we show (columns 2 and 4 respectively) the mean position of the ionization front $\langle R_{\rm IF} \rangle$ at several times and the standard deviation, $\sigma$, as obtained by the six different 1D codes. We find that the position $r$ is remarkably similar between the codes, as the relative difference is $\lesssim1\%$ in all cases, and that this difference decreases with time. In the same table we show (in column 6) the mean maximum density reached, $\langle \rho_{\rm max} \rangle$, as well as the respective standard deviation (in column 7). As before, better agreement is obtained as $t$ increases.

At later times, the H~{\sc ii} region expansion slows down monotonically, and so in reality the shock Mach number (hence compression factor and maximum density) also decreases monotonically. The shell mass and thickness also increase monotonically with time as more mass is swept up. This means that the quoted $\langle\rho_{\rm max}\rangle$ is certainly an underestimate compared to the analytic solution at early times, and gradually reaches the correct value at later times. This can be seen in the first panel of Fig.~\ref{fig:1Dearly}, where all of the codes have different peak densities because of the differing spatial resolution. The largest density achieved (by {\sc glide}) is significantly larger than the mean value quoted in Table \ref{tab:early}.

\begin{table}
\caption{ This table shows results from the `early phase' test (see \ref{sssec:1Dearly}). Column 1 shows the time. Columns 2 and 4 show the mean position of the ionization front, $\langle R_{\rm IF} \rangle$ with the corresponding standard deviation, $\sigma$, for 1D codes. Columns 3 and 5 are the respective for 3D codes. Columns 6 and 7 show the mean maximum density of the shock front $\langle\rho_{\rm max}\rangle$ and its standard deviation, $\sigma$, for 1D codes only.}
\centering
\label{tab:early}
\begin{tabular}{c c c c c c c}
\hline
$t$ & \multicolumn{2}{c}{$\langle R_{\rm IF} \rangle$} & \multicolumn{2}{c}{$\sigma/10^{-3}$} & $\langle\rho_{\rm max}\rangle/10^{-19}$& $\sigma/10^{-19}$ \\
(Myr) & \multicolumn{2}{c}{(pc)} & \multicolumn{2}{c}{(pc)}  & ($\rm g\,cm^{-3}$) & ($\rm g\,cm^{-3}$)\\
\hline
 & 1D & 3D & 1D & 3D & 1D & 1D \\
\hline
$0.005$  &  $0.373$ &  $0.378$  &   $2.059$ & $2.76$ & $5.7$ & $3.6$\\
$0.01$   &  $0.430$ &  $0.430$  &   $3.180$ & $2.96$ & $5.9$ & $3.0$\\
$0.02$   &  $0.536$ &  $0.530$  &   $2.488$ & $3.26$ & $6.1$ & $3.3$\\
$0.04$   &  $0.717$ &  $0.707$  &   $3.848$ & $3.59$ & $4.2$ & $1.4$\\
$0.08$   &  $1.009$ &  $0.996$  &   $3.679$ & $4.35$ & $2.6$ & $0.4$\\
$0.14$   &  $1.343$ &  $1.324$  &   $2.848$ & $4.29$ & $1.7$ & $0.2$\\
\hline
\end{tabular}
\end{table}

\subsubsection{Late phase test (Test 2-1D)}
\label{sssec:1Dlate}

\begin{figure*}
\centering
\includegraphics[width=0.45\textwidth]{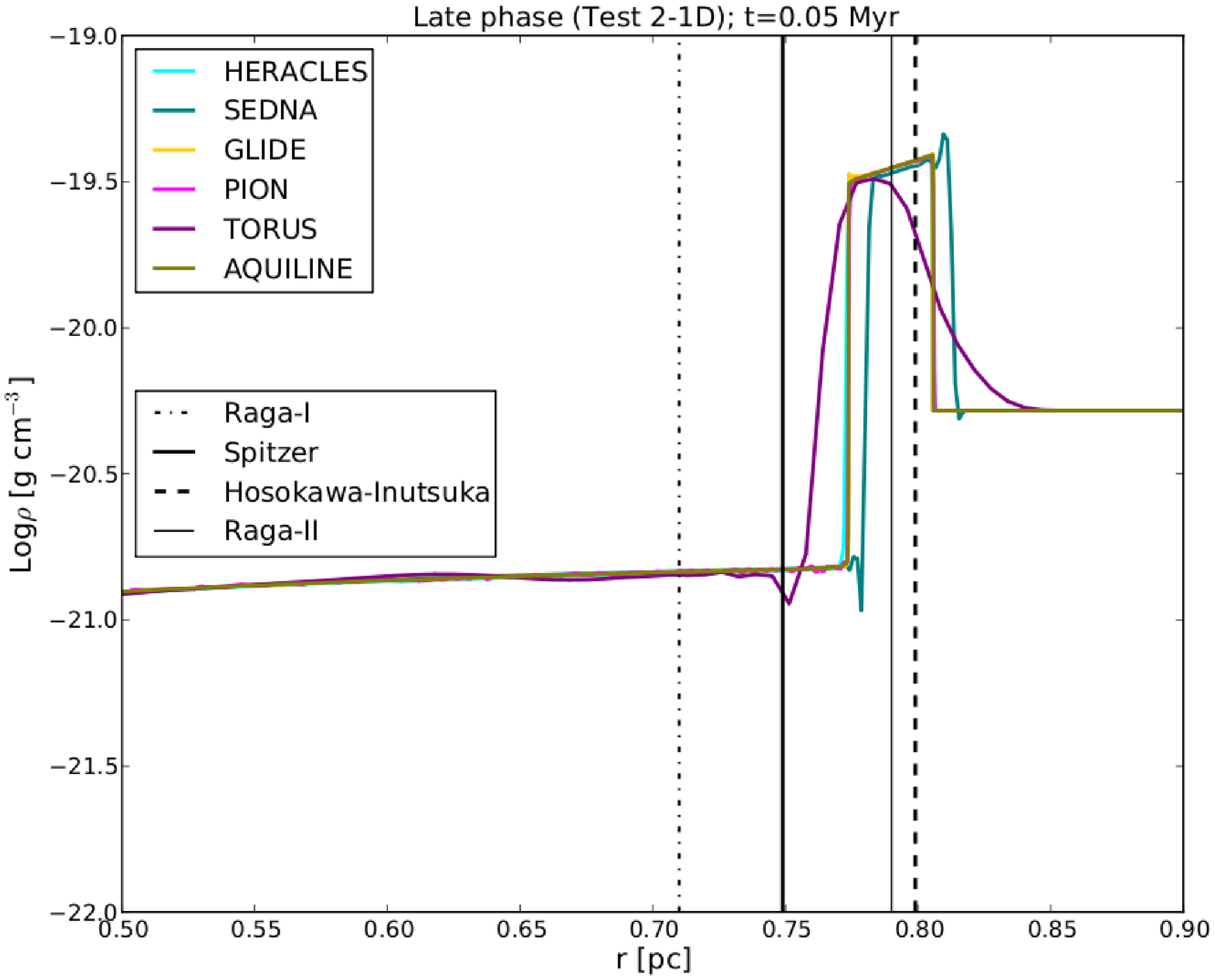}
\includegraphics[width=0.45\textwidth]{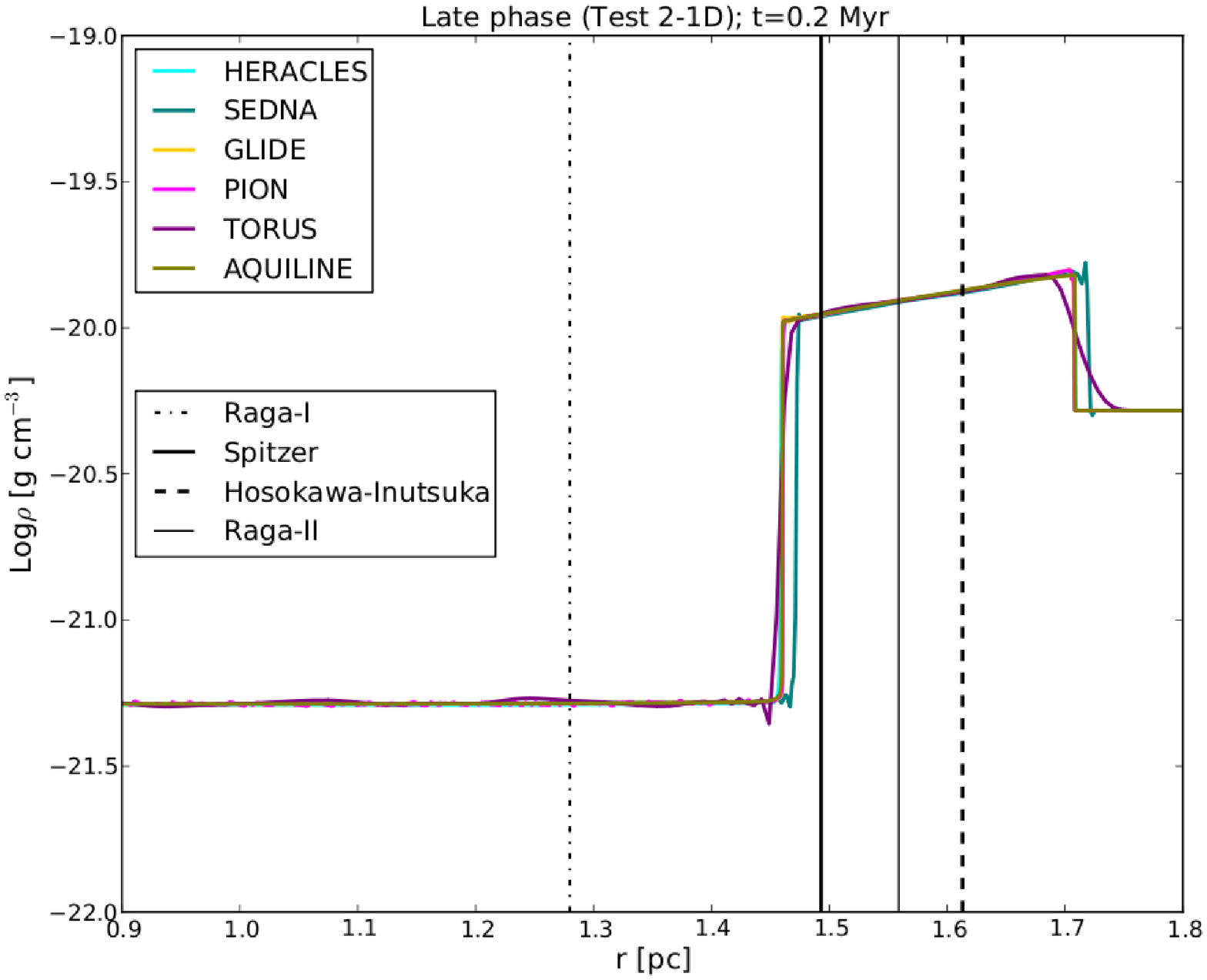}
\includegraphics[width=0.45\textwidth]{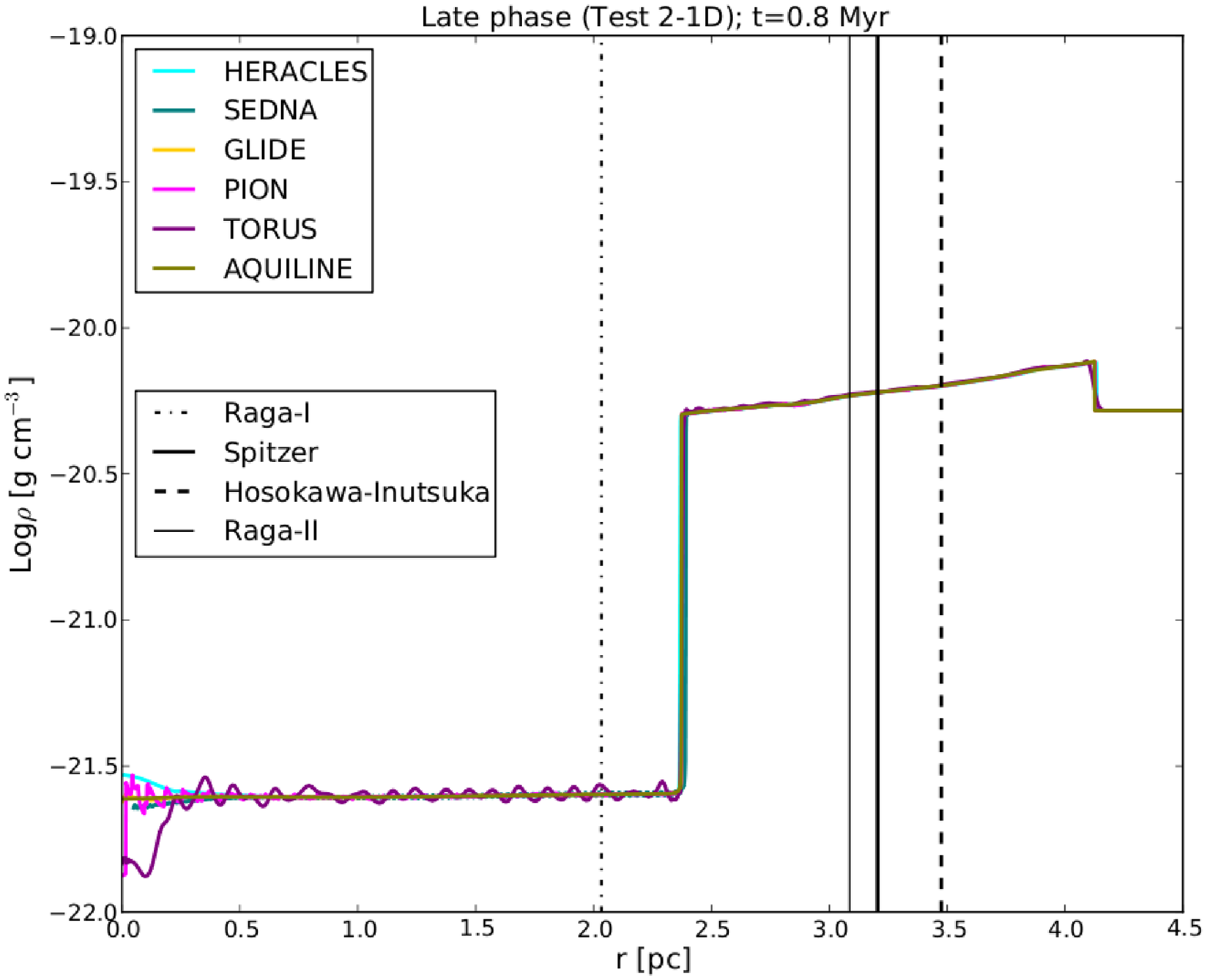}
\includegraphics[width=0.45\textwidth]{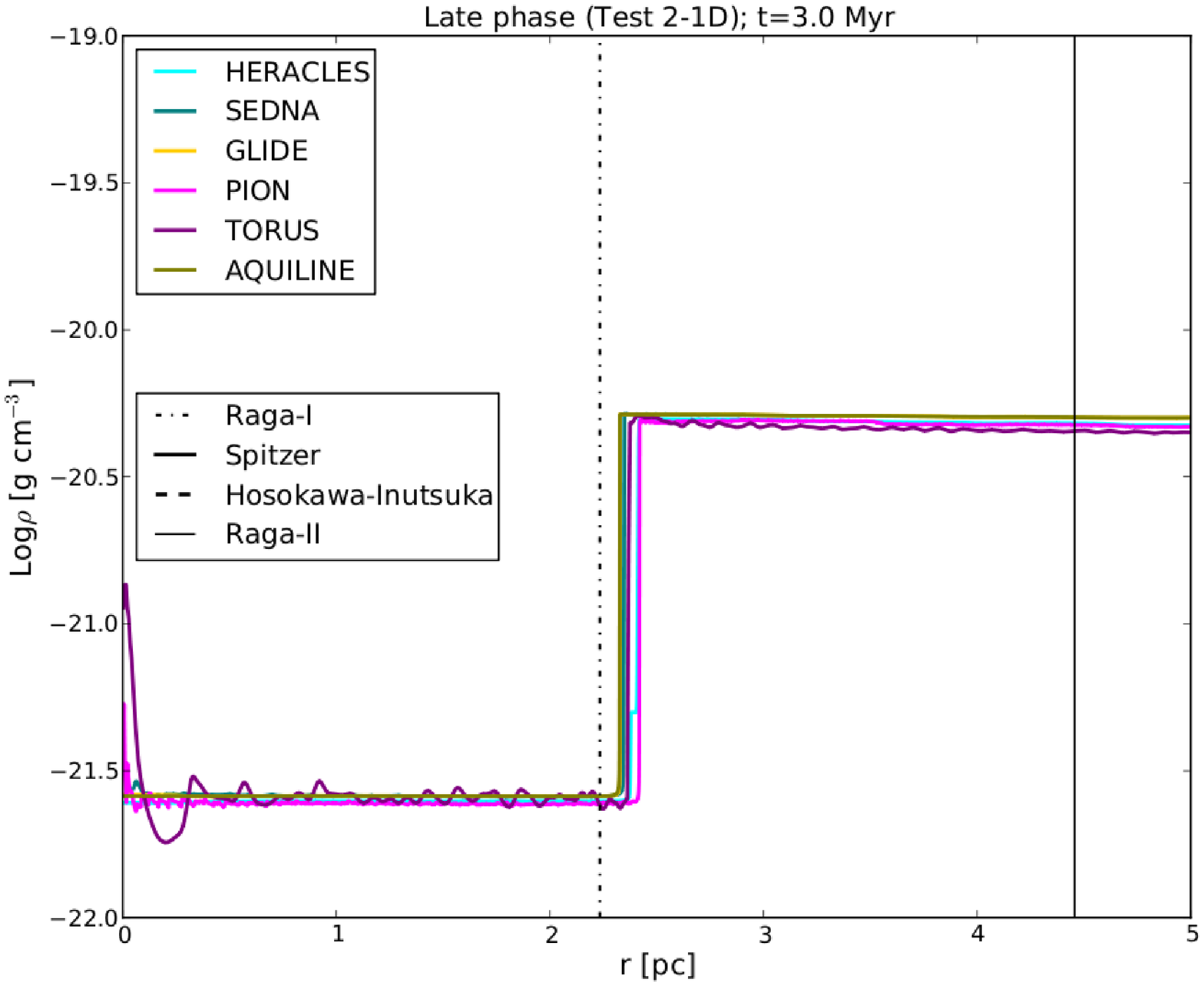}
\caption{ As in Fig.~\ref{fig:1Dearly} for the late phase test. We plot snapshots at $t=0.05\,{\rm Myr}$ (top left), $t=0.2\,{\rm Myr}$ (top right), $t=0.8\,{\rm Myr}$ (middle left), and $t=3.0\,{\rm Myr}$ (middle right).}
\label{fig:1Dlate}
\end{figure*}

Figure \ref{fig:1Dlate} shows the density as a function of radius for the six participating 1D codes in the Test 2-1D for different times, i.e. $t=0.05$, $t=0.2$, $t=0.8$ and $t=3.0\,{\rm Myr}$. The vertical lines correspond to the positions predicted by the analytical expressions discussed in Section \ref{sec:theory}. Table \ref{tab:late} gives the mean position of the ionization front with the standard deviation (columns 2 and 4) as well as the mean and standard deviation for the maximum density reached (columns 6 and 7). The relative difference around $\langle R_{\rm IF} \rangle$ is $\lesssim2\%$ in all cases. 

In this late phase test, the position of the ionization front stagnates as predicted by Eqn.~\ref{eqn.raga} (Raga-I) since the H~{\sc ii} region is in pressure equilibrium with the pressure of the neutral medium. Indeed, as shown in Table \ref{tab:late}, the mean stagnation radius (measured at $t=3.0\,{\rm Myr}$) as given by all six participating codes is $\langle R_{\rm IF} \rangle=2.359\,{\rm pc}$ (standard deviation $\sigma=0.032\,{\rm pc}$ or $\sim1.3\%$ from this mean value) which gives a relative error of $\sim2.2\%$ when compared to Eqn.~\ref{eqn.rstag}. 

The shell thickness is a function of resolution, which is why the thickness obtained by {\sc torus} is larger than the other results (i.e. 1024 cells, compared to a maximum of 40960 used in {\sc pion}). The shell develops rapidly and becomes geometrically thick by $t=0.05\,{\rm Myr}$. As seen in Fig.~\ref{fig:1Dlate}, the thickness is remarkably similar as calculated by all six participating codes. The shell becomes thicker over time as observed for $t\ge0.2\,{\rm Myr}$.  The leading shock eventually becomes a detached expanding compression wave, and the value of the maximum density $\rho_{\rm max}$ drops down towards $\sim\rho_{\rm o}$ which is the density of the undisturbed neutral medium (i.e. at the beginning of each simulation). 

\begin{figure}
\centering
\includegraphics[width=0.45\textwidth]{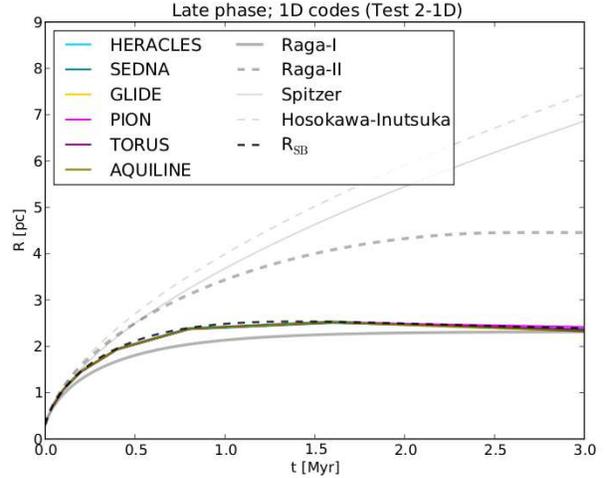}
\caption{ Comparison between Eqns.~\ref{eqn.raga} (Raga-I), and \ref{eqn.raga2diff} (Raga-II) for the 1D participating codes. The black dashed line corresponds to the {\sc StarBench} Eqn.~\ref{eqn:rsb}. We additionally plot the Spitzer and the Hosokawa-Inutsuka (HI) expansion laws for reference.}
\label{fig:1Dlaterrors}
\end{figure}

Figure \ref{fig:1Dlaterrors} shows the position of the ionization front of all codes compared with Eqns.\ref{eqn.raga} (Raga-I) and \ref{eqn.raga2diff} (Raga-II). For reference, we also plot the Spitzer and Hosokawa-Inutsuka equations. As first shown by \citet{Raga12b}, the codes initially follow the position predicted by Eqn.~\ref{eqn.raga2diff} however they `relax' towards the stagnation radius expected from pressure balance predicted by Eqn.~\ref{eqn.rstag}. The position of the ionization front temporarily expands beyond the stagnation radius before relaxing inwards. According to \citet{Raga12b}, this `overshoot' is due to the inertia of the H~{\sc ii} region.

Here, we provide a semi-empirical equation (which we call the `{\sc StarBench} equation') predicting the position of the ionization front for this late phase test for $t\in[0,3]\,{\rm Myr}$. The {\sc StarBench} equation has the form:
\begin{eqnarray}
\label{eqn:rsb}
R_{\rm SB}={\cal R}_{\rm II}+f_{\rm SB}({\cal R}_{\rm I}-{\cal R}_{\rm II}),
\end{eqnarray}
where $\cal{R}_{\rm I}$ and $\cal{R}_{\rm II}$ are the numerical solutions of Eqns.\ref{eqn.raga} and \ref{eqn.raga2diff} respectively, and $f_{\rm SB}$ is assumed to be a time-dependent dimensionless factor. For the purposes of the present paper we approximate $f_{\rm SB}$ with the following form
\begin{eqnarray}
\label{eqn:fsb}
f_{\rm SB}=1-{\rm A}\exp\left[-\frac{t}{\rm Myr}\right],
\end{eqnarray}
where ${\rm A}=0.733$. This equation is plotted in Fig.~\ref{fig:1Dlaterrors} and \ref{fig:3Dlaterrors} and shows very good agreement with all 1D and 3D simulations. We propose that coders who benchmark their algorithm against this late phase test should use Eqns.\ref{eqn.raga}, \ref{eqn.raga2diff}, \ref{eqn:rsb}, and \ref{eqn:fsb} with the indicated constant for $t\gtrsim0.05\,{\rm Myr}$.

\begin{table}
\caption{ As in Table~\ref{tab:early} for the `late phase' test.}
\centering
\label{tab:late}
\begin{tabular}{c c c c c c c}
\hline
$t$ & \multicolumn{2}{c}{$\langle R_{\rm IF} \rangle$} & \multicolumn{2}{c}{$\sigma/10^{-3}$} & $\langle\rho_{\rm max}\rangle/10^{-20}$& $\sigma/10^{-21}$ \\
(Myr) & \multicolumn{2}{c}{(pc)} & \multicolumn{2}{c}{(pc)}  & ($\rm g\,cm^{-3}$) & ($\rm g\,cm^{-3}$)\\
\hline
 & 1D & 3D & 1D & 3D & 1D & 1D \\
\hline
 $0.05$ &  $0.771$ &  $0.733$  &  $8.685$  & $2.606$ &  $3.90$ & $4.0$ \\
 $0.1$  &  $1.063$ &  $1.025$  &  $4.356$  & $2.741$ &  $2.43$ & $1.8$ \\
 $0.2$  &  $1.460$ &  $1.420$  &  $6.772$  & $2.645$ &  $1.57$ & $0.5$ \\
 $0.4$  &  $1.934$ &  --       &  $6.366$  & --      &  $1.05$ & $0.4$ \\
 $0.8$  &  $2.375$ &  $2.349$  &  $7.450$  & $6.739$ &  $0.73$ & $0.8$ \\
 $1.6$  &  $2.515$ &  $2.506$  &  $10.07$  & $5.854$ &  $0.58$ & $0.6$ \\
 $3.0$  &  $2.359$ &  $2.424$  &  $31.71$  & $1.534$ &  $0.53$ & $0.3$ \\
\hline
\end{tabular}
\end{table}

\subsubsection{Summary of 1D results}

Overall we find good agreement between all six 1D participating codes. In order to achieve this we require such a high resolution that a 3D calculation would be prohibitively expensive \citep[$\gg 10^4$ cells in each dimension for a uniform grid, see also][for the relevant discussion in SPH]{Bisb09}. Although all codes are in excellent agreement at sufficiently fine mesh scales, we find limited agreement with the existing and widely used analytic forms. We attribute this to the approximations made in deriving these analytic forms. Many of these approximations are only valid when applied to the early phase thin shell expansion. We provide a tuned fitting formula for which all codes mutually agree to $\lesssim0.5\%$ error at all times.

\subsection{3D runs}
\label{ssec:r3d}

\subsubsection{Early phase test (Test 1-3D)}
\label{sssec:3Dearly}

\begin{figure}
\centering
\includegraphics[width=0.45\textwidth]{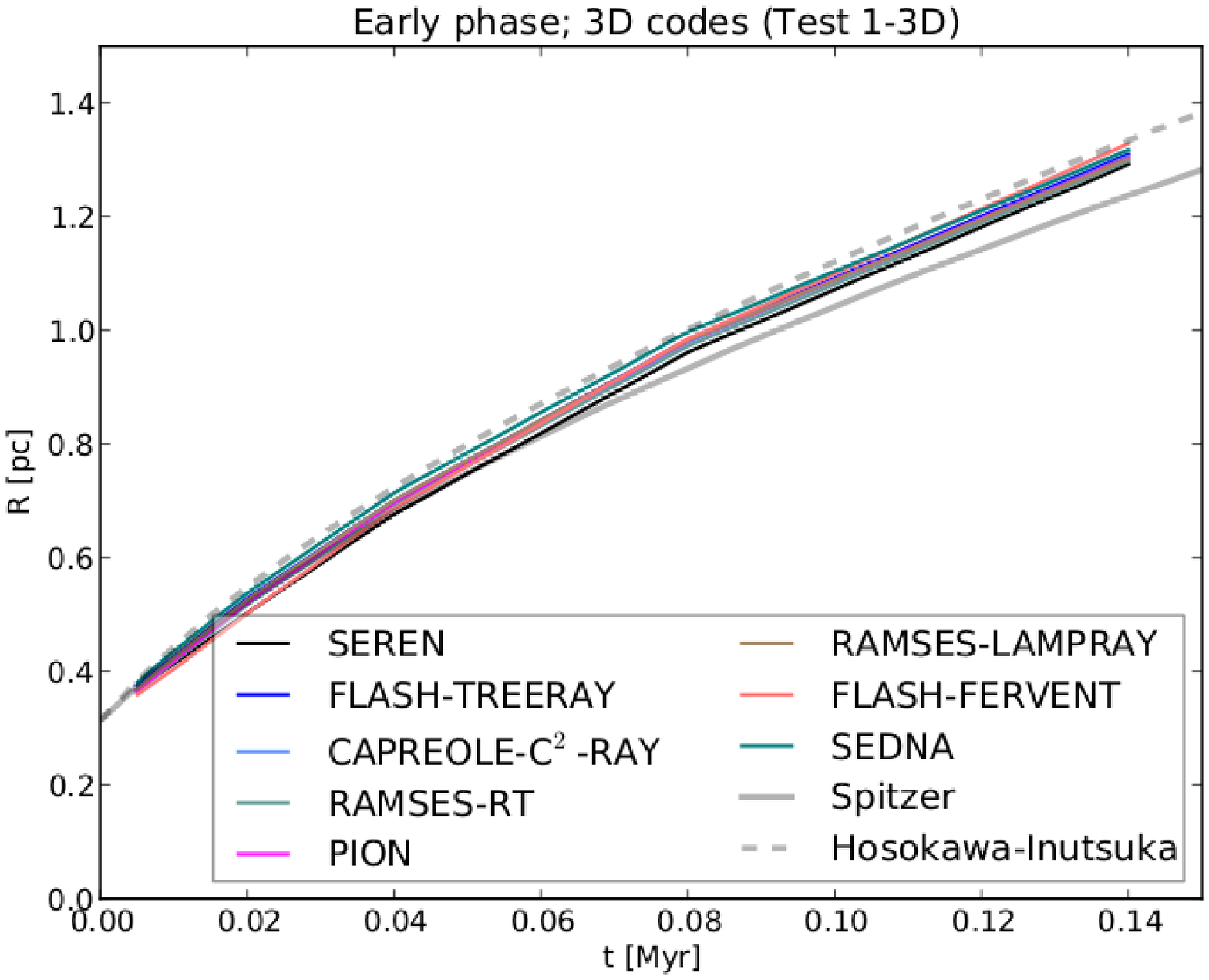} 
\includegraphics[width=0.45\textwidth]{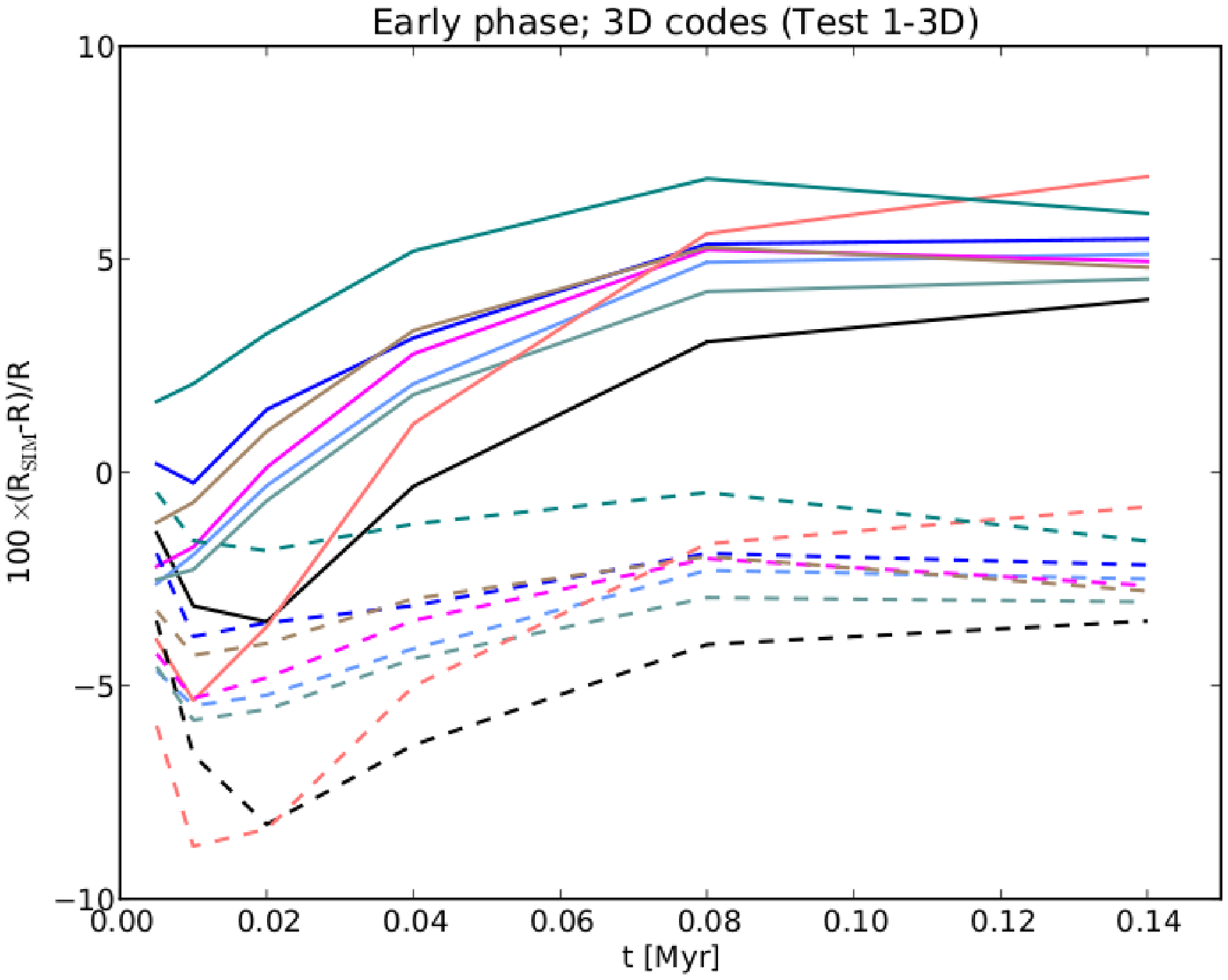}  
\caption{ As in Fig.~\ref{fig:1Dearlyerrors} for the 3D participating codes.}
\label{fig:3Dearlyerrors}
\end{figure}

In Appendix \ref{app} we present (Figs. \ref{fig:X3Dlate_a} and \ref{fig:X3Dlate_b}) snapshots of slices through the density distributions at various times during the evolution of the H~{\sc ii} region for the eight 3D participating codes. In general, the agreement in structure and front position between the widely different codes is excellent.

There are, however, some significant differences. At $t=0.005\,{\rm Myr}$, the peak density of the shell as simulated by the SPH code {\sc seren} is significantly lower than the other codes and the shell is broader. However, the agreement improves at later times. If the number of SPH particles is increased, this leads to a better agreement suggesting that a higher particle count is required to achieve comparable resolution to the grid-based codes. The number of SPH particles used was tuned to match the overall expansion of the H~{\sc ii} region (i.e. $10^4$ particles for the initial Str{\"o}mgren sphere) rather than to resolve the shell structure. However, the total particle count ($6.4\times10^5$) is comparable to the number of zones in the Eulerian calculations ($128^3$).

As time progresses, evidence of instability appears in some of the calculations. In particular, the instabilities are particularly prominent in {\sc flash-treeray} and {\sc capreole-C$^2$-Ray}, which both use the PPM algorithm to steepen discontinuities. It should be mentioned here that in these particular cases the instabilities develop first in the places where the shock in the neutral gas is propagating parallel to the grid and that if a different cut through the 3D datacube is taken, the instabilities are not in evidence. This strongly suggests that for these two codes the instability is of the odd-even type, first described by \citet{Quir94}. These stabilities can be cured by introducing extra numerical diffusion into the PPM scheme when a strong shock is detected parallel to the grid direction, or using a hybrid Riemann solver, which switches from the usual Riemann solver (e.g. the Roe solver used by {\sc capreole-C$^2$-Ray}) to a more diffusive Riemann solver such as an HLLE solver, inside shocks.

Figure \ref{fig:3Dearlyerrors} shows the comparison of the analytical equations, as we discussed above, to the eight participating 3D codes. Here, we find that all codes also follow the expansion law as expressed by the Hosokawa-Inutsuka Eqn.~\ref{eqn:HI}, with error $\lesssim4\%$ at $t\gtrsim0.08\,{\rm Myr}$ in contrast to $\gtrsim5\%$ when compared to the Spitzer Eqn.~\ref{eqn.Spitzer}. Table \ref{tab:early} shows the mean position, $\langle R_{\rm IF} \rangle$ (column 3), of the ionization front and the standard deviation (column 5) for all these codes which we find to agree to within $3-7\%$ around $\langle R_{\rm IF} \rangle$. As noted earlier, all codes agree with the Spitzer approximation only for $t\lesssim0.07\,{\rm Myr}$, however based on the 1D simulations, we argue that this value of time will be decreased down to $\sim0.01\,{\rm Myr}$ by increasing the resolution substantially.

\subsubsection{Late phase test (Test 2-3D)}
\label{sssec:3Dlate}

\begin{figure}
\centering
\includegraphics[width=0.45\textwidth]{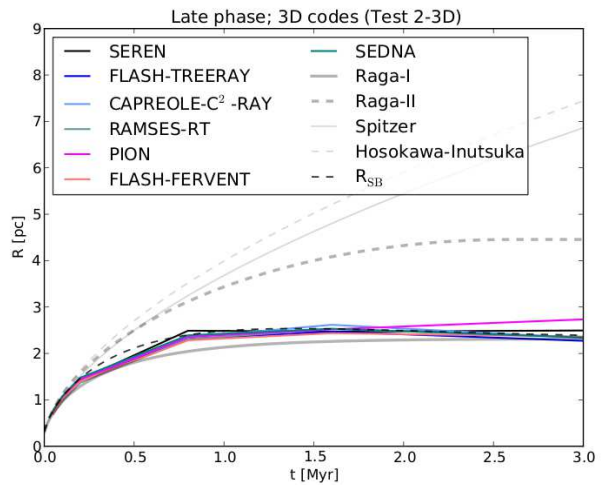}
\caption{ As in Fig.~\ref{fig:1Dlaterrors} for the 3D participating codes.}
\label{fig:3Dlaterrors}
\end{figure}

Figures \ref{fig:X3Dlate_a} and \ref{fig:X3Dlate_b} show cross-section density plots at times $t=0.05$, $t=0.2$, $t=0.8$, and $t=3.0\,{\rm Myr}$. We find that all codes are in excellent qualitative agreement until $t\sim0.8\,{\rm Myr}$. From $t\approx0.8\,{\rm Myr}$, most of the codes show some level of D-type instability similar to that shown in \citet{Will02}. This has the largest magnitude in {\sc pion}, for which the resulting Reynolds stresses appear to be causing the ionization front to expand to a larger radius than found in the other codes. 

In the case of the SPH code {\sc seren}, the shell becomes prone to the tensile instability \citep{Mona00} and it creates high density contrast fluctuations in its interior. Eventually at $t=3.0\,{\rm Myr}$ the shell has been completely detached and the hot ionized medium is bounded by vacuum. One would expect that this ionized region would expand rapidly into the external vacuum as long as the shell is detached, however it appears that because there are no SPH particles in this region the SPH summations do not sample it and hence the edge of the region functions as a smooth flow boundary condition. This may be a similar issue to the known contact discontinuity problem in SPH \citep{Ager07, Pric08}.

Figure \ref{fig:3Dlaterrors} shows a comparison of the position of the ionization front as calculated by all participating 3D codes and the analytical solutions. We additionaly plot the {\sc StarBench} Eqn.~\ref{eqn:rsb}. As we explained in \S\ref{sssec:3Dearly}, none of the codes reproduce any of the analytical solutions. Table \ref{tab:late} shows the mean values of the distance of the ionization front and the corresponding standard deviation. All codes are in very good mutual agreement within $\lesssim5\%$.

\section{Summary \& Conclusions}
\label{sec:sumcon}%

We have studied the standard radiation hydrodynamical test case of an H~{\sc ii} region expanding in an initially uniform density medium. Twelve distinct codes participated in this {\sc StarBench} comparison test. We examined two scenarios of the initial `early phase' expansion and the `late phase' relaxation to pressure equilibrium with the external medium. 

The early phase test shows that the Hosokawa-Inutsuka approximation (Eqn.~\ref{eqn:HI}) which results directly from the equation of motion of the expanding shell and which is in fact a second order ODE, agrees with the numerical results. In contrast, the Spitzer approximation (Eqn.~\ref{eqn.Spitzer}) which results from the assumption that the thermal pressure of the H~{\sc ii} region matches approximately with the ram pressure, and which is a first order ODE, underestimates the numerical results by a small-but-significant factor ($\sim8\%$), because it does not include the effect of the inertia of the material entrained in the shell.

In the late phase test, we tested the codes against Eqns.\ref{eqn.raga} (Raga-I) and \ref{eqn.raga2diff} (Raga-II) which include the pressure of the undisturbed medium acting on the H~{\sc ii} region and are generalizations of Eqns.\ref{eqn.Spitzer} (Spitzer) and \ref{eqn:HI} (Hosokawa-Inutsuka) respectively. With time, this pressure becomes approximately equal to the thermal pressure within the H~{\sc ii} region, at which point the system reaches equilibrium and the expansion of the ionization front is halted. Our benchmarking test showed that all participating codes start by following the expansion law as expressed by Eqn.~\ref{eqn.raga2diff} (Raga-II), while at later times the H~{\sc ii} region stagnates at the position predicted by static pressure balance.

Since the codes do not agree with any of the analytic expressions, we have developed an analytic parametrization (which we call the `StarBench equation') that describes the early and late phase expansion of an H~{\sc ii} region to within $\lesssim2\%$ at all times which we recommend be used for future code validations, exercises and analytical studies of massive star-forming regions. 

The structure of the expanding H~{\sc ii} region is overall in very good agreement between all the contributing codes. We have discussed physics-related issues, such as instabilities, which apply in certain case and which may either be physical or numerical. This agreement between the codes has improved dramatically as result of the comparison test we present in this paper. This comparison has increased the confidence of the scientific robustness of the results of all participating codes.

\section*{Acknowledgements}

The authors thank the referee Tom Hartquist for the comments.
The work of TGBisbas was funded by STFC grant ST/J001511/1. 
The work of TJHaworth was funded by STFC grant ST/K000985/1.
The work of RJRWilliams contains material (c) British Crown Owned Copyright 2015/AWE.
JMackey acknowledges support from the Deutsche Forschungsgemeinschaft priority program 1573, Physics of the Interstellar Medium. 
The work of PTremblin is partly supported by the European Research Council under the European Community’s Seventh Framework Programme (FP7/20072013 Grant Agreement No. 247060). 
SJArthur acknowledges support from DGAPA-UNAM through project IN101713.
RKuiper acknowledges funding from the Max Planck Research Group Star formation throughout the Milky Way Galaxy at the Max Planck Institute for Astronomy. RKuiper further acknowledges funding from the Emmy Noether Research Group on ``Accretion Flows and Feedback in Realistic Models of Massive Star Formation'' granted by the German Research Foundation under grant no. KU 2849/3-1.
SGeen is funded by European Research Council under the European Community's Seventh Framework Programme (FP7/2007-2013) / ERC Grant Agreement (no. 306483).
ITIliev acknowledges support by STFC grants ST/F002858/1 and ST/I000976/1.
JRosdahl is funded by the European Research Council under the European Unions Seventh Framework Programme (FP7/2007-2013) / ERC Grant agreement 278594-GasAroundGalaxies, and the Marie Curie Training Network CosmoComp (PITN-GA-2009-238356).
TFrostholm and THaugboelle are supported by the Danish National Research Foundation, through its establishment of the Centre for Star and Planet Formation, and by a Sapere Aude Starting Grant to THaugboelle from the Danish Council for Independent Research. The work by THaugboelle used the astrophysics cluster at the University of Copenhagen HPC facility, supported by a research grant (VKR023406) from VILLUM FONDEN.
SWalch acknowledges support by the DFG through SPP 1573 and through SFB 956.
RW\"unsch acknowledges support by the project RVO:67985815 and by grant 209/12/1795 of the Czech Science Foundation.

Part of this work used the DiRAC Data Analytics system at the University of Cambridge, operated by the University of Cambridge High Performance Computing Serve on behalf of the STFC DiRAC HPC Facility (www.dirac.ac.uk). This equipment was funded by BIS National E-infrastructure capital grant (ST/K001590/1), STFC capital grants ST/H008861/1 and ST/H00887X/1, and STFC DiRAC Operations grant ST/K00333X/1. DiRAC is part of the National E-Infrastructure. Part of this work used the Large Facilities Capital Fund of BIS and the University of Oxford supercomputing facilities. 

TGBisbas, TJHaworth, RJRWilliams, JMackey, PTremblin, ACRaga, SJArthur, CBaczynski, JEDale and ITIliev acknowledge the NORDITA programme on Photo-Evaporation in Astrophysical Systems (2013 June) where significant part of the work for this paper was carried out. 

The second {\sc StarBench} workshop (Bonn, September 2014) was supported by a grant from the Deutsche Forschungsgemeinschaft priority program 1573, Physics of the Interstellar Medium. This research has made use of NASA's Astrophysics Data System.

\appendix

\section{Further plots}
\label{app}

\begin{figure}
\centering
\includegraphics[width=0.48\textwidth]{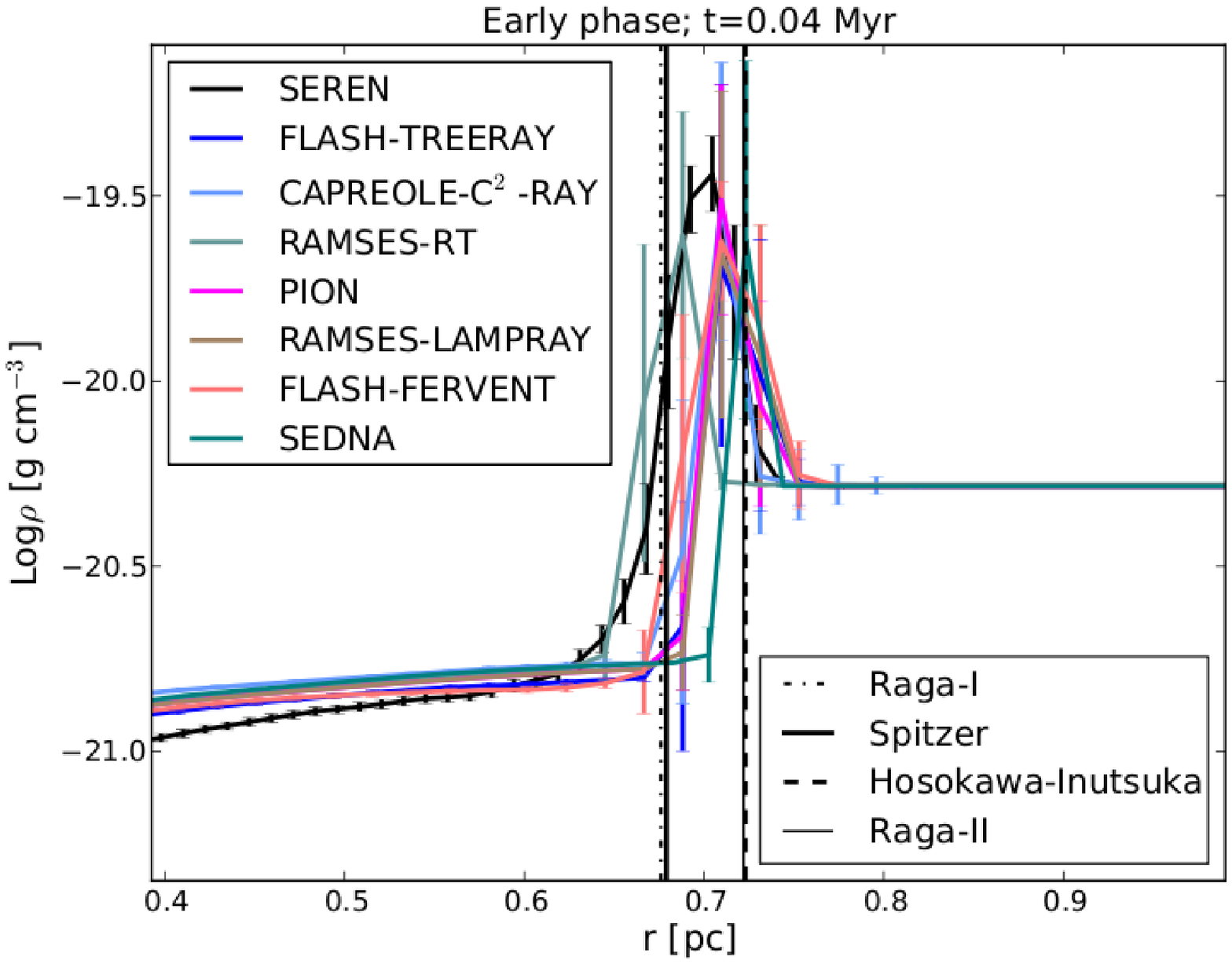}
\includegraphics[width=0.48\textwidth]{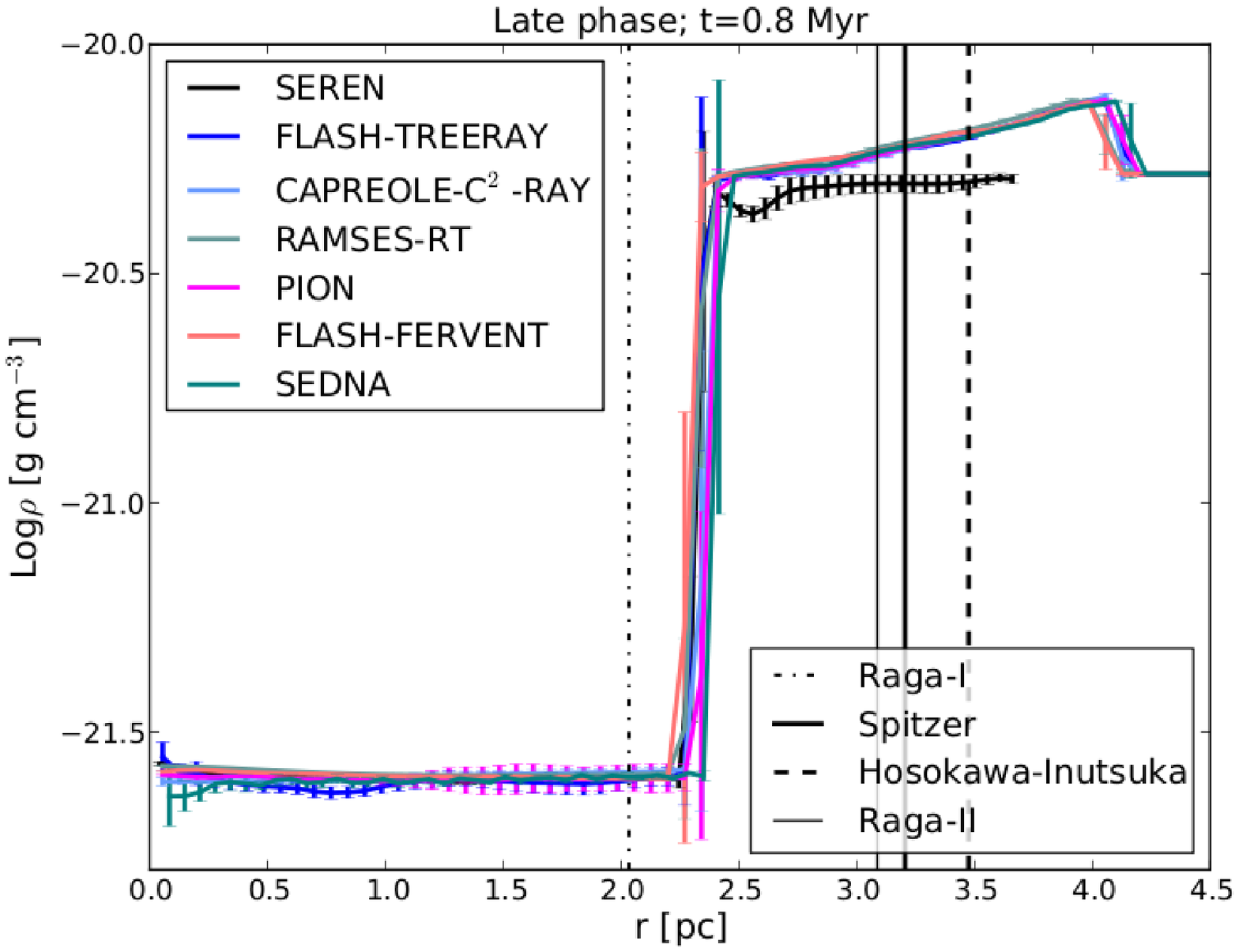}
\caption{ Density plots of the early phase test (Test 1-3D; left) and the late phase test (Test 2-3D; right) for all 3D participating codes. The $x$-axis is in pc and the $y$-axis in $\rho\,{\rm cm}^{-3}$. The bars indicate $1\sigma$.}
\label{fig:3Ddensity_a}
\end{figure}

Figure~\ref{fig:3Ddensity_a} shows density plots of the 3D codes for the `early phase' test (Test 1-3D; left panel) and the `late phase' test (Test 2-3D; right panel). Both panels are similar to the plots of Figs.~\ref{fig:1Dearly} and \ref{fig:1Dlate}. The bars correspond to $1\sigma$ standard deviation from all cells/SPH particles.

\begin{figure*}
\centering
\includegraphics[width=0.95\textwidth]{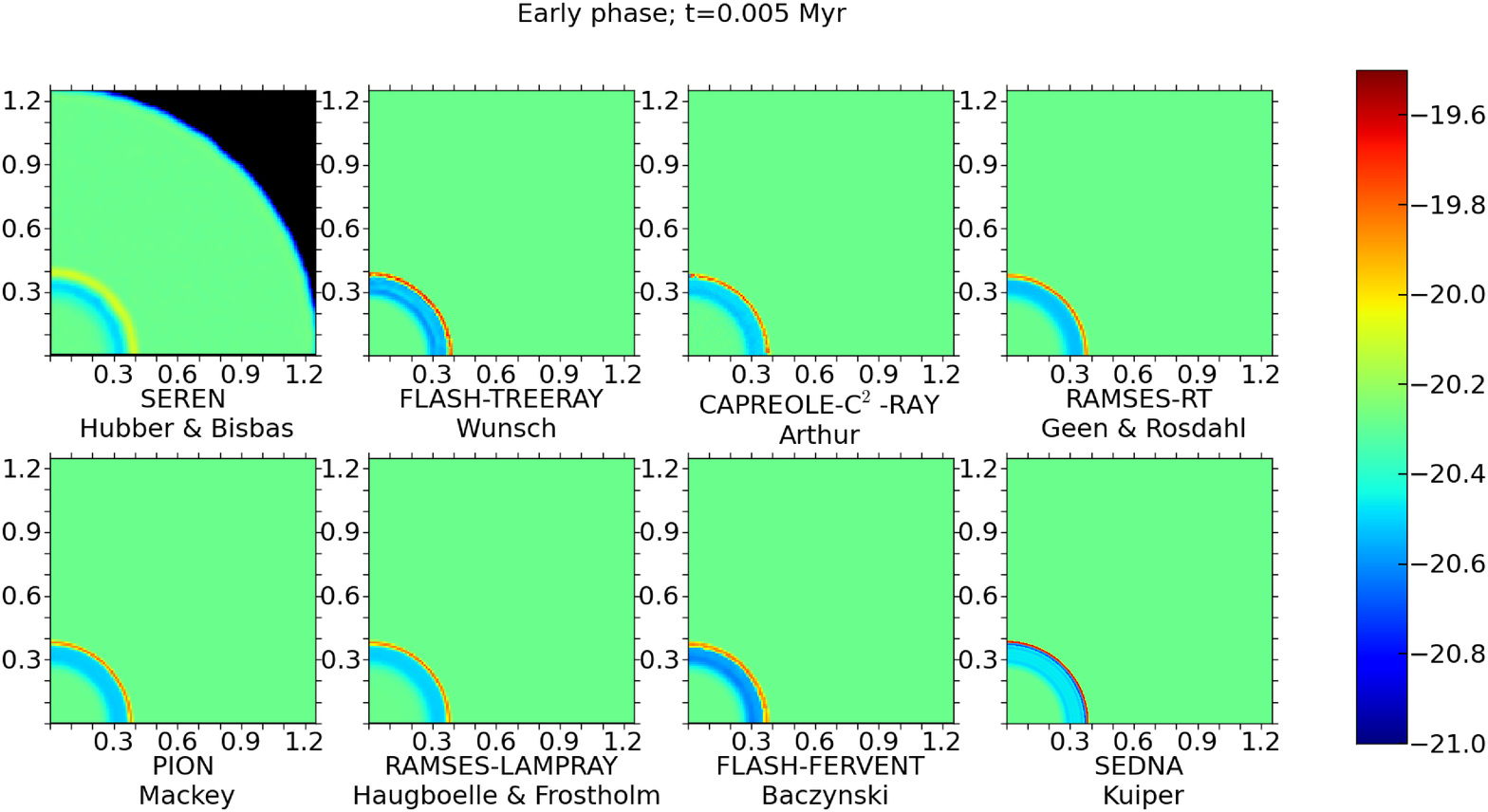}
\includegraphics[width=0.95\textwidth]{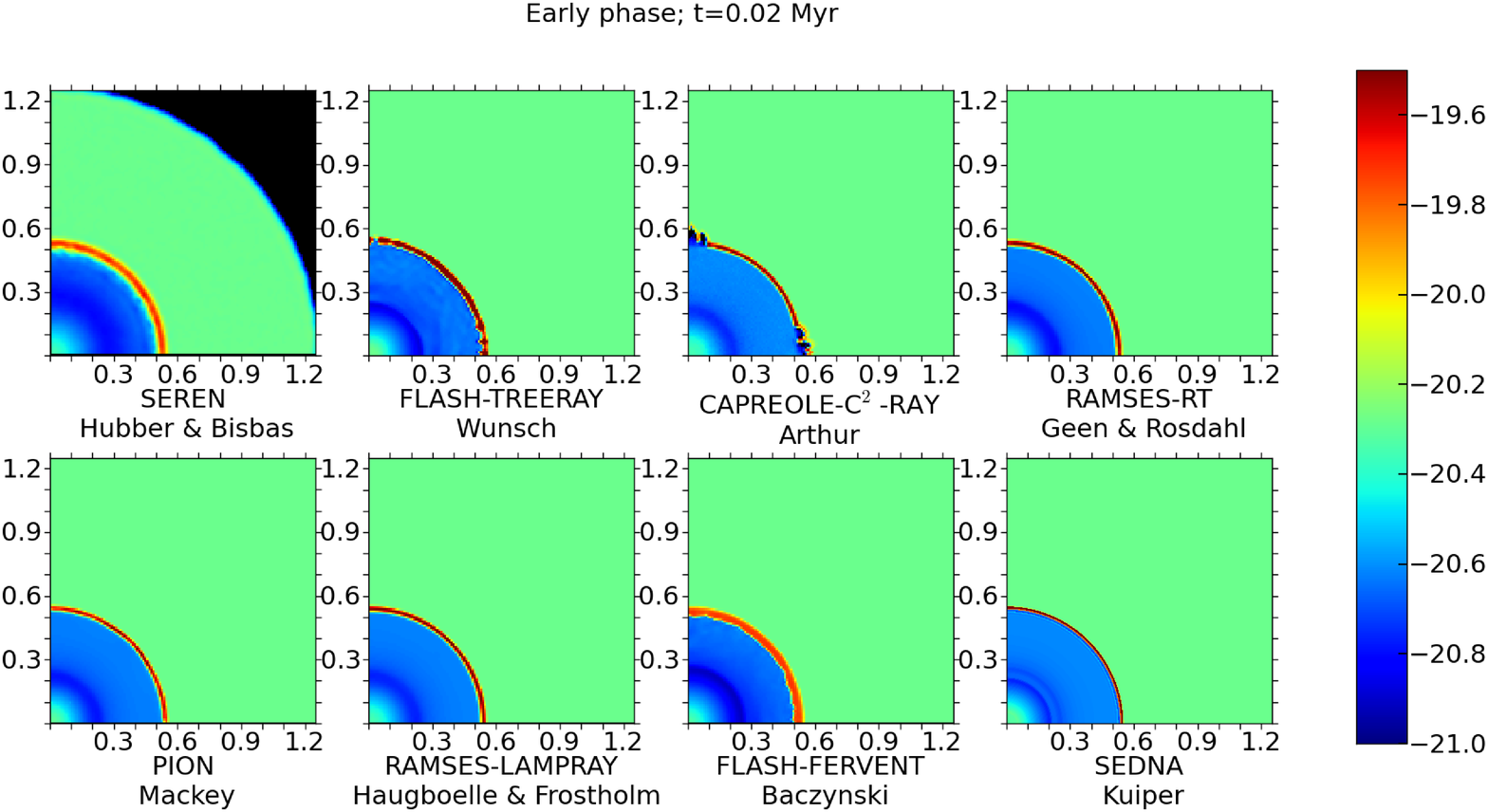}
\caption{ Cross section (slice) plots taken at $z=0\,{\rm pc}$ for the early phase (Test 1-3D) expansion of all participating 3D codes described in \S\ref{sec:codes}. $x$- and $y$- axes are in pc. The logarithmic colour bar shows gas density in ${\rm g}\,{\rm cm}^{-3}$. We show snapshots at $t=0.005\,{\rm Myr}$ and $t=0.02\,{\rm Myr}$.}
\label{fig:X3Dearly_a}
\end{figure*}

\begin{figure*}
\centering
\includegraphics[width=0.95\textwidth]{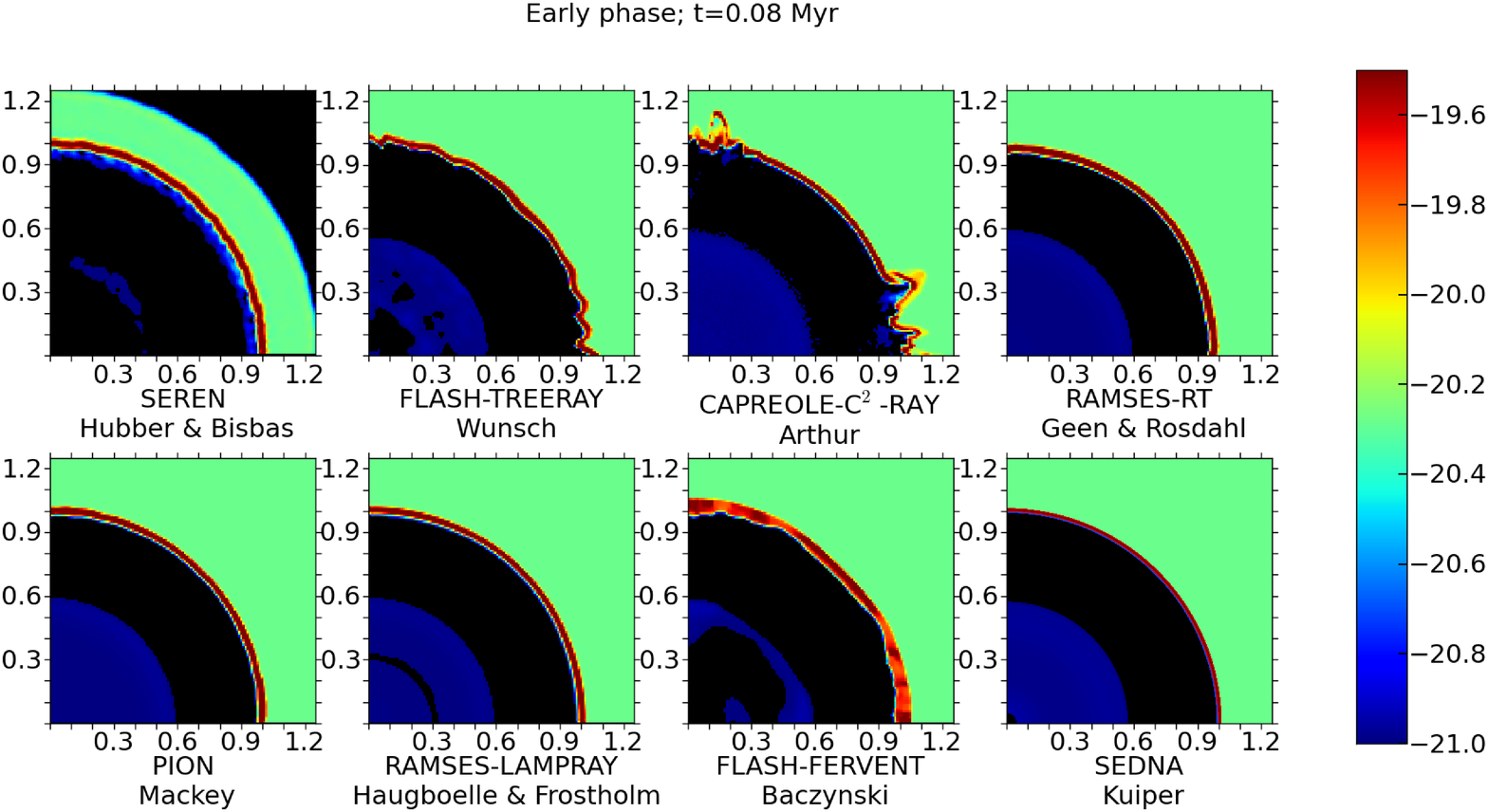}
\includegraphics[width=0.95\textwidth]{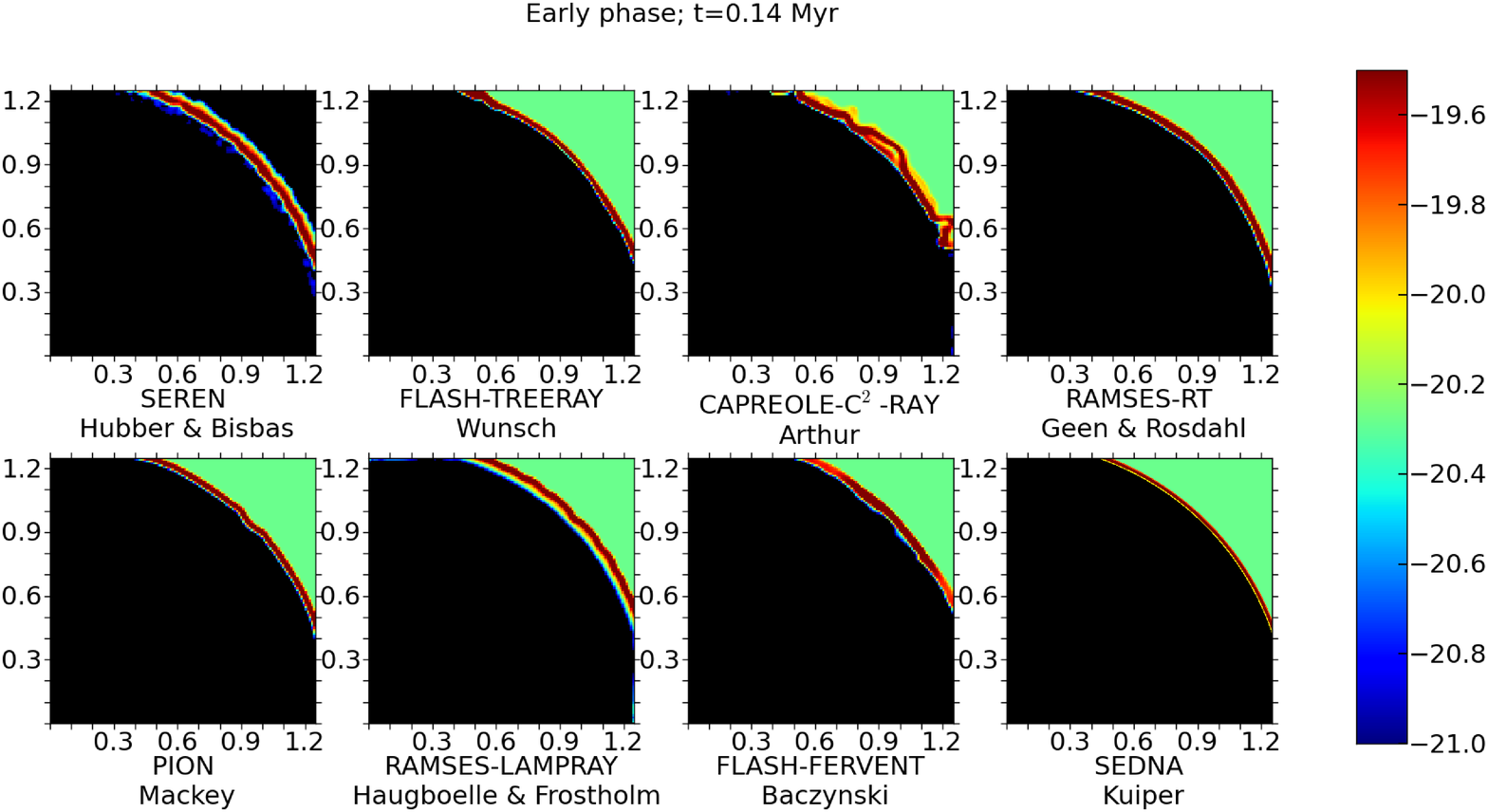}
\caption{ As in Fig.~\ref{fig:X3Dearly_a} for $t=0.08\,{\rm Myr}$ and $t=0.14\,{\rm Myr}$.}
\label{fig:X3Dearly_b}
\end{figure*}

\begin{figure*}
\centering
\includegraphics[width=0.95\textwidth]{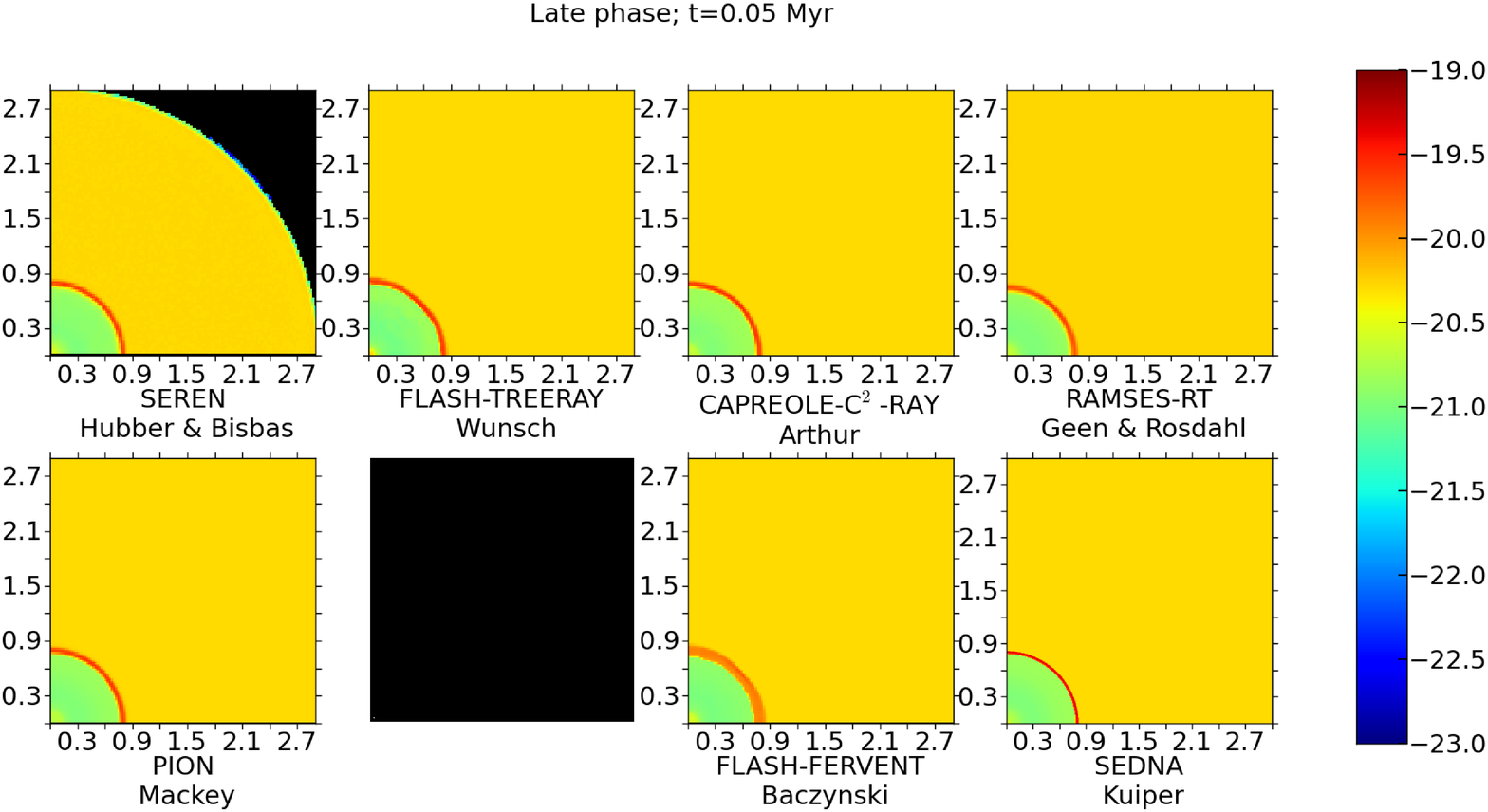}
\includegraphics[width=0.95\textwidth]{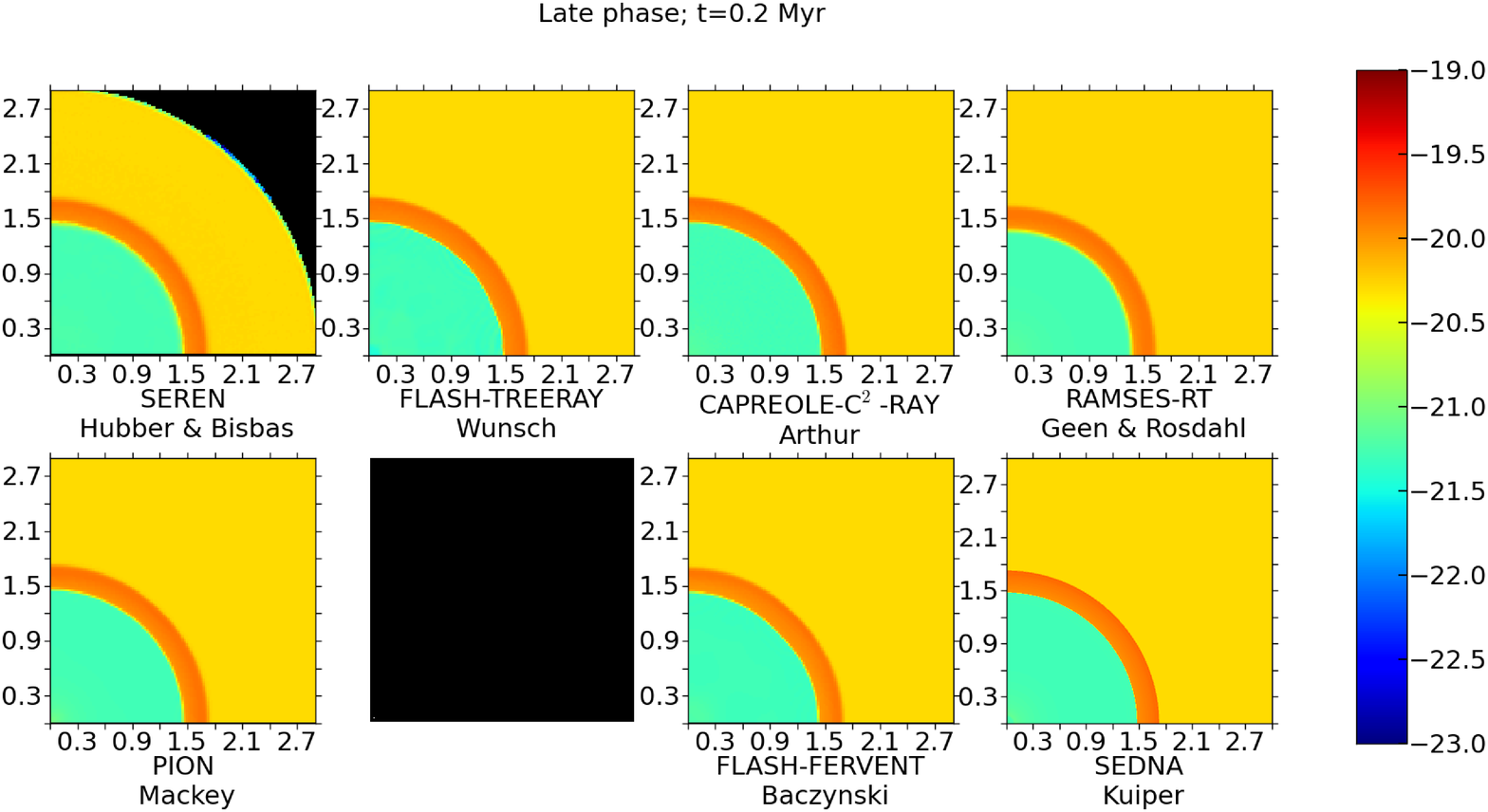}
\caption{ Cross section (slice) plots taken at $z=0\,{\rm pc}$ for the late phase (Test 2-3D) expansion of all participating 3D codes described in \S\ref{sec:codes}. $x$- and $y$-axes are in pc. The logarithmic colour bar shows gas density in g\,cm$^{-3}$. We show snapshots at $t=0.05\,{\rm Myr}$ and $t=0.2\,{\rm Myr}$.}
\label{fig:X3Dlate_a}
\end{figure*}

\begin{figure*}
\centering
\includegraphics[width=0.95\textwidth]{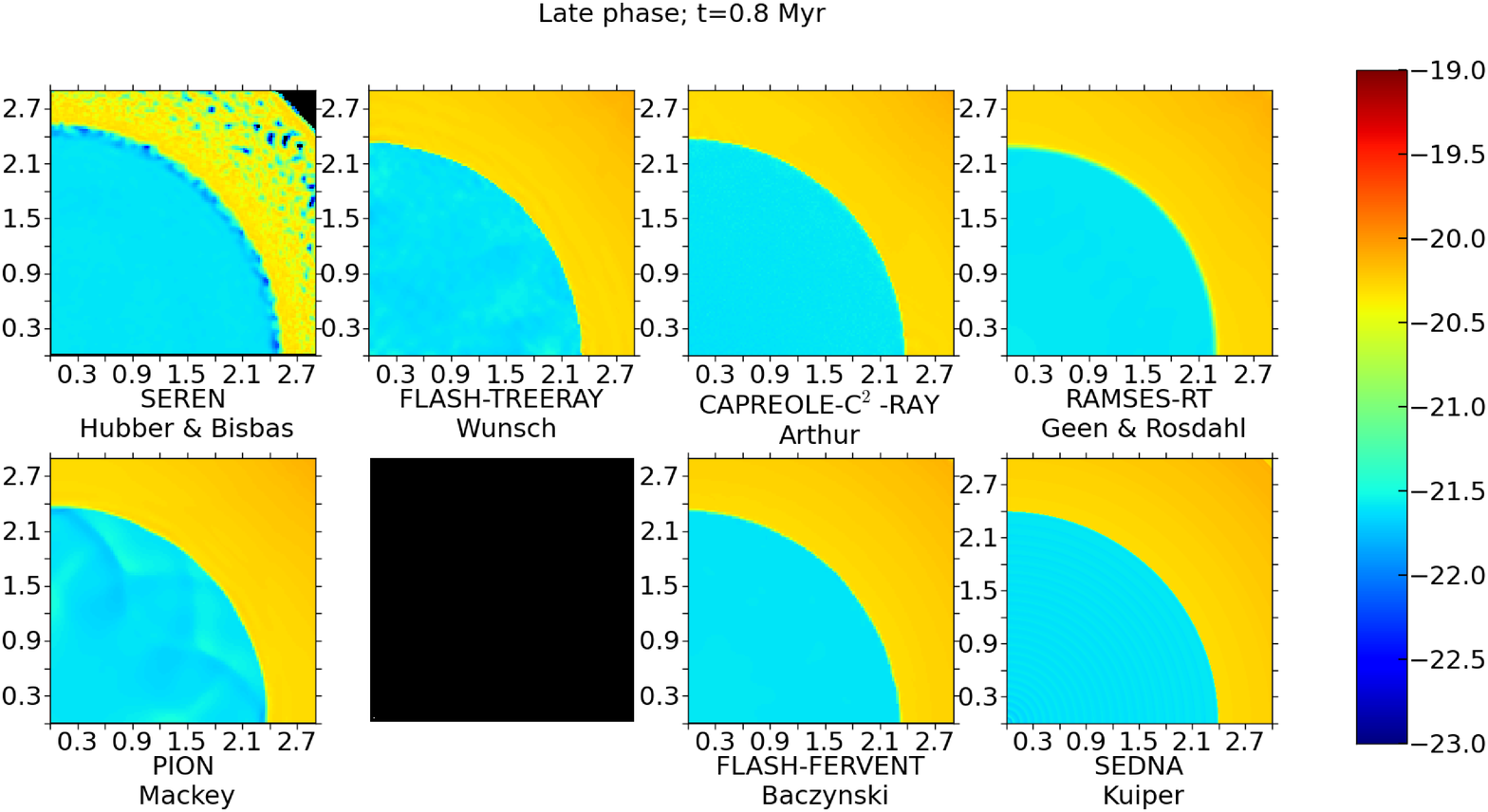}
\includegraphics[width=0.95\textwidth]{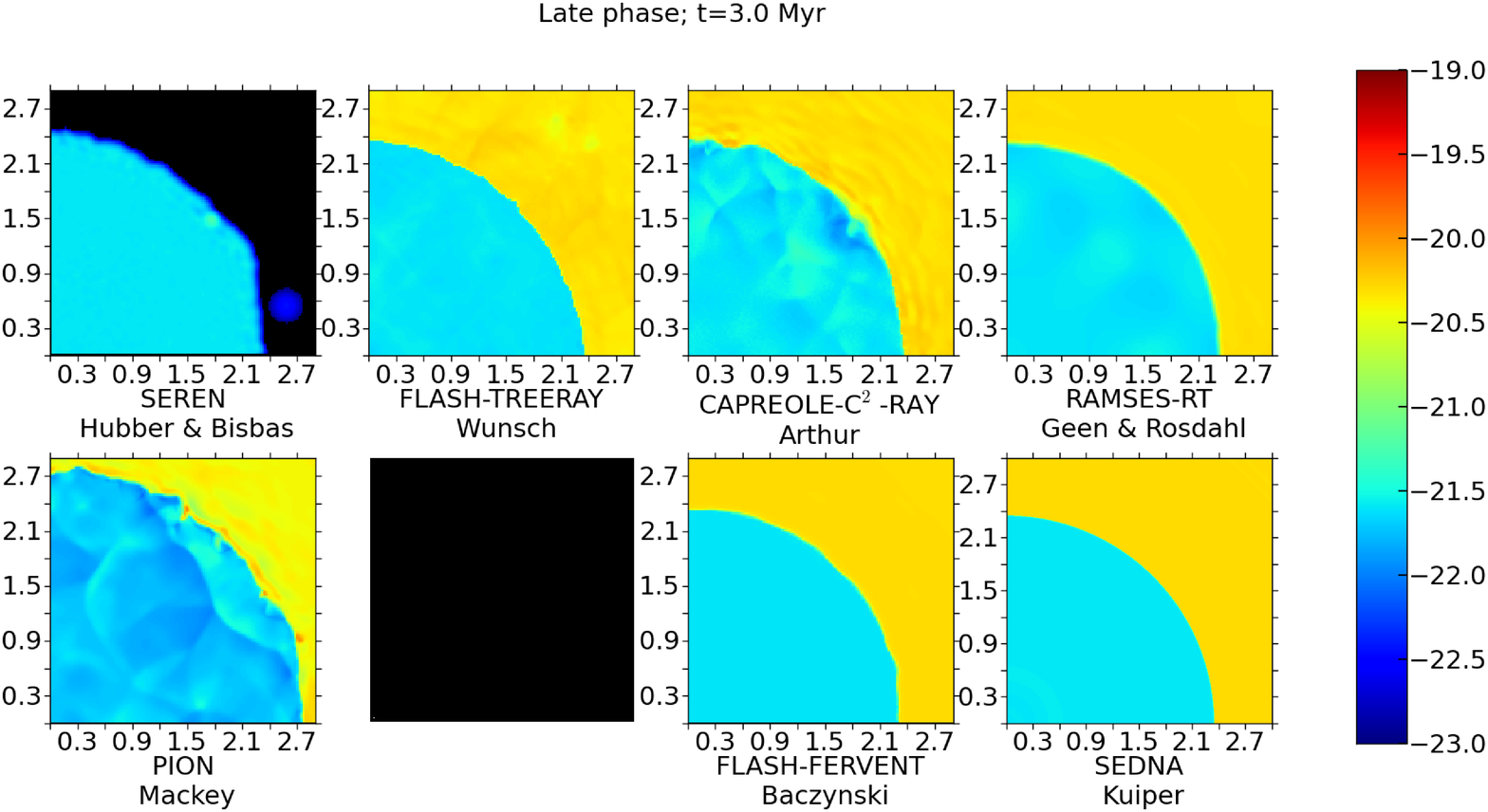}
\caption{ As in Fig.~\ref{fig:X3Dlate_a} but for $t=0.8\,{\rm Myr}$ and $t=3.0\,{\rm Myr}$.}
\label{fig:X3Dlate_b}
\end{figure*}

Figures~\ref{fig:X3Dearly_a}-\ref{fig:X3Dlate_b} show cross section (slice) plots of all 3D codes for the `early' and `late' phase tests at $z=0\,{\rm pc}$. All 3D grid codes have used the same spatial resolution (128 cells) and mesh geometry (uniform). For the particular case of the SPH code {\sc seren}, we have remapped the SPH particles on the $z=0\,{\rm pc}$ slice and have assumed a $128^2$ resolution \citep[see][for further details]{Pric07}.

\bsp

\label{lastpage}

\end{document}